\pdfoutput=1
\documentclass[fleqn,usenatbib]{mnras}

\usepackage{newtxtext,newtxmath}

\usepackage[T1]{fontenc}
\usepackage{ae,aecompl}

\usepackage{graphicx}	
\usepackage{amsmath}	
\usepackage{amssymb}	

\usepackage{multicol}
\usepackage{xspace}
\usepackage{xcolor}
\usepackage{threeparttable}
\usepackage{tablefootnote}
\usepackage{caption}
\usepackage{subcaption}
\usepackage{printlen}
\usepackage{comment}

\makeatletter
\def\@to{to}
\makeatother
\definecolor{cadmiumgreen}{rgb}{0.0, 0.42, 0.24}

\title[Anatomy of the Orphan Stream]{SMHASH: Anatomy of the Orphan Stream using RR Lyrae stars}

\author[D. Hendel et al.]{David~Hendel$^1$\thanks{E-mail: hendel@astro.columbia.edu},
Victoria~Scowcroft$^{2,3}$, Kathryn~V.~Johnston$^1$, Mark~A.~Fardal$^4$, 
\newauthor Roeland~P.~van~der~Marel$^{4,5}$, Sangmo~Tony~Sohn$^{4,5}$, Adrian~M.~Price-Whelan$^6$,  
\newauthor Rachael~L.~Beaton$^6$, Gurtina~Besla$^{7}$,  Giuseppe~Bono$^{8,9}$, Maria-Rosa~L.~Cioni$^{10,11}$, 
\newauthor Gisella~Clementini$^{12}$, Judith~G.~Cohen$^{13}$, Michele~Fabrizio$^{14,9}$, Wendy~L.~Freedman$^{15}$, 
\newauthor Alessia~Garofalo$^{16}$, Carl~J.~Grillmair$^{17}$,  Nitya~Kallivayalil$^{18}$, Juna~A.~Kollmeier$^3$,  
\newauthor David~R.~Law$^4$, Barry~F.~Madore$^3$, Steven~R.~Majewski$^{19}$, Massimo~Marengo$^{20}$, 
\newauthor Andrew~J.~Monson$^3$, Jillian~R.~Neeley$^{20}$, David~L.~Nidever$^{21}$, Grzegorz~Pietrzy\'{n}ski$^{22}$, 
\newauthor Mark~Seibert$^3$, Branimir~Sesar$^{23}$, Horace~A.~Smith$^{24}$, Igor~Soszy\'{n}ski$^{25}$,
\newauthor Ian~B.~Thompson$^3$, and Andrezej~Udalski$^{25}$\\
$^1$Department of Astronomy, Columbia University, 550 West 120th Street, New York, NY 10027, USA\\
$^2$Department of Physics, University of Bath, Claverton Down, Bath, BA2 7AY, UK \\
$^3$Observatories of the Carnegie Institution of Washington, 813 Santa Barbara St., Pasadena, CA 91101, USA\\
$^4$Space Telescope Science Institute, 3700 San Martin Drive, Baltimore, MD 21218, USA\\
$^5$Department of Physics and Astronomy, The Johns Hopkins University, Baltimore, MD 21218, USA\\
$^6$Department of Astrophysical Sciences, Princeton University, 4 Ivy Lane, Princeton, NJ~08544\\
$^{7}$Steward Observatory, University of Arizona, 933 North Cherry Avenue, Tucson, AZ 85721, USA\\
}

\date{Accepted XXX. Received YYY; in original form ZZZ}

\pubyear{2017}

\begin{document}
\label{firstpage}
\pagerange{\pageref{firstpage}--\pageref{lastpage}}
\maketitle

\begin{abstract}
Stellar tidal streams provide an opportunity to study the motion and structure of the disrupting galaxy as well as the gravitational potential of its host. Streams around the Milky Way are especially promising as phase space positions of individual stars will be measured by ongoing or upcoming surveys. Nevertheless, it remains a challenge to accurately assess distances to stars farther than 10 kpc from the Sun, where we have the poorest knowledge of the Galaxy's mass distribution. To address this we present observations of 32 candidate RR Lyrae stars in the Orphan tidal stream taken as part of the {\it Spitzer} Merger History and Shape of the Galactic Halo (SMHASH) program. The extremely tight correlation between the periods, luminosities, and metallicities of RR Lyrae variable stars in the {\it Spitzer} IRAC  $\mathrm{3.6 \mu m}$ band allows the determination of precise distances to individual stars; the median statistical distance uncertainty to each RR Lyrae star is $2.5\%$. By fitting orbits in an example potential we obtain an upper limit on the mass of the Milky Way interior to 60 kpc of $\mathrm{5.6_{-1.1}^{+1.2}\times 10^{11}\ M_\odot}$, bringing estimates based on the Orphan Stream in line with those using other tracers. The SMHASH data also resolve the stream in line--of--sight depth, allowing a new perspective on the internal structure of the disrupted dwarf galaxy. Comparing with N--body models we find that the progenitor had an initial dark halo mass of approximately $\mathrm{3.2 \times 10^{9}\ M_\odot}$, placing the Orphan Stream's progenitor amongst the classical dwarf spheroidals.
\end{abstract}


\section{Introduction}

Tidal debris structures are striking evidence of hierarchical assembly -- the premise that the Milky Way and systems like it have been built over cosmic time through the coalescence of many smaller objects \citep{1978MNRAS.183..341W, 1996ApJ...465..278J, 2001ApJ...548...33B, 2002ARA&A..40..487F}. Some of this construction is in the form of major mergers, where two near--equal mass galaxies collide and their stars are redistributed wholesale as the new galaxy violently relaxes. However, in the prevailing $\Lambda$ -- cold dark matter ($\Lambda$CDM) model, the vast majority of mergers (by number) are minor \citep{2010MNRAS.406.2267F} where one halo, the host, dominates the interaction and a smaller object, the satellite, is dragged inward by dynamical friction and eventually stripped of mass by tidal forces. When the luminous component is disrupted the stars may form a stellar tidal stream or shell, depending on the parameters of the interaction \citep[e.g.][]{2008ApJ...689..936J, 2015MNRAS.450..575A, 2015MNRAS.454.2472H}. The study of tidal features therefore probes the accretion histories of galaxies.

\footnotetext[8]{Dipartimento di Fisica, Universit\`{a} di Roma Tor Vergata, via Della Ricerca Scientifica 1, I-00133, Roma, Italy}
\footnotetext[9]{INAF, Rome Astronomical Observatory, via Frascati 33, I-00040, Monte Porzio Catone, Italy}
\footnotetext[10]{Leibniz-Instit\"{u}t f\"{u} Astrophysik Potsdam, An der Sternwarte 16, 14482 Potsdam, Germany}
\footnotetext[11]{University of Hertfordshire, Physics Astronomy and Mathematics, College Lane, Hatfield AL10 9AB, UK}
\footnotetext[12]{INAF-Osservatorio Astronomico di Bologna, Via Ranzani 1, I - 40127 Bologna, Italy}
\footnotetext[13]{Division of Physics, Mathematics and Astronomy, California Institute of Technology, Pasadena, Ca., 91125}
\footnotetext[14]{Space Science Data Center - ASI, via del Politecnico, s.n.c., I-00133, Roma, Italy}
\footnotetext[15]{Department of Astronomy \& Astrophysics, University of Chicago, 5640 South Ellis Avenue, Chicago, IL 60637, USA}
\footnotetext[16]{Dipartimento di Fisica e Astronomia, Universit\`{a} di Bologna,Viale Berti Pichat 6/2, I-40127 Bologna, Italy}
\footnotetext[17]{IPAC, Mail Code 314-6, Caltech, 1200 E. California Blvd., Pasadena, CA 91125}
\footnotetext[18]{Department of Astronomy, University of Virginia, 530 McCormick Road, Charlottesville, VA 22904, USA}
\footnotetext[19]{Department of Astronomy, University of Virginia, P.O. Box 400325, Charlottesville, VA 22904-4325, USA}
\footnotetext[20]{Department of Physics and Astronomy, Iowa State University, A313E Zaffarano, Ames, IA 50010}
\footnotetext[21]{National Optical Astronomy Observatory, 950 North Cherry Avenue, Tucson, AZ 85719, USA}
\footnotetext[22]{Nicolaus Copernicus Astronomical Center, Polish Academy of Sciences, Bartycka 18, 00-716 Warsaw, Poland}
\footnotetext[23]{Max Planck Institute for Astronomy, Konigstuhl 17, D-69117, Heidelberg, Germany}
\footnotetext[24]{Department of Physics and Astronomy, Michigan State University, East Lansing, MI 48824-2320, USA}
\footnotetext[25]{Warsaw University Observatory, Al. Ujazdowskie 4, 00-478 Warszawa, Poland}

Stellar tidal streams are also key tools for our current understanding of the Milky Way's gravitational potential. The techniques applied to measure the potential are wide--ranging but commonly a few--parameter potential model is varied in an attempt to match simulations to the available data. Historically, the streams used most often for this purpose are the Sagittarius dwarf galaxy's stream \citep{2003ApJ...599.1082M, 2010ApJ...714..229L,2014MNRAS.445.3788G} and various globular cluster streams such as Palomar~5 and GD-1 \citep{2010ApJ...712..260K, 2015ApJ...803...80K, 2015ApJ...799...28P, 2015ApJ...811..123F, 2016ApJ...833...31B}.

The Orphan tidal stream \citep{2006ApJ...645L..37G,2006ApJ...642L.137B} has several advantages over the other streams mentioned above. It forms a smooth arc that is significantly longer \citep[detected length of $\approx 108^{\circ},$][]{2015ApJ...812L..26G}, wider \citep[$\sim 2^{\circ}$,][]{2006ApJ...642L.137B}, and farther from the Galactic centre \citep[$> 50$ kpc,][]{2010ApJ...711...32N,2013ApJ...776...26S} than any of the commonly studied globular cluster streams. Along with its total luminosity \citep[$\mathrm{M_r < -7.5,}$][]{2007ApJ...658..337B} and metallicity spread of $0.56$ dex \citep{2013ApJ...764...39C}, these characteristics suggest a dwarf spheroidal galaxy as the likely origin, but the progenitor is elusive and possibly nearly completely disrupted by the Galaxy's tidal field \citep{2015ApJ...812L..26G}. In contrast to the Sagittarius stream, the Orphan Stream has a uniform appearance and cold velocity structure; the Sagittarius stream is notoriously complex, featuring multiple wraps, bifurcated tails, and several stellar populations with different kinematics \citep{2006ApJ...642L.137B, 2012ApJ...750...80K, 2017MNRAS.464..794G}. The orbital planes of the Orphan and Sagittarius streams are misaligned by $\sim 67^{\circ}$ \citep{2012MNRAS.423.1109P}, making the combination of the two an attractive target for multi--stream potential measuring methods \citep{2015ApJ...801...98S, 2016ApJ...833...31B}.

The Orphan Stream also has the advantage of a well--filled horizontal branch resulting in numerous classes of stars that may be used as standard candles for distance estimation, for example the Blue Horizontal Branch (BHB) stars studied by \cite{2010ApJ...711...32N}. Of particular relevance to this work, the Orphan Stream contains a significant population of RR Lyrae stars (RRL), which have been the focus of several recent efforts to improve distance measurement into the Galactic halo \citep{2017ApJ...844L...4S,2017arXiv171009436H}. These stars make excellent standard candles using period--luminosity (PL) relations with their near-- or mid--infrared magnitudes \citep{1986MNRAS.220..279L, 2001MNRAS.326.1183B, 2003MNRAS.344.1097B, 2004ApJS..154..633C, 2015ApJ...799..165B}. In addition to the advantage of decreased extinction at these longer wavelengths compared to the V band ($A_{\mathrm{V}}/A_{\mathrm{[3.6\mu m]}} > 19$, \citealt{1989ApJ...345..245C, 2005ApJ...619..931I}), the PL relation has also been shown to have a small intrinsic scatter in the infrared \citep{2013ApJ...776..135M,2015ApJ...808...11N}. Recently these relations have been extended to include a metallicity component \citep{2017ApJ...841...84N} with the effect of further decreasing the uncertainty on individual stars' absolute magnitudes and thus removing systematic scatter in measured distances for systems with a large range in metallicity, such as the Orphan stream. 

The {\it Spitzer} Merger History And Shape of the Galactic Halo (SMHASH) program builds upon the previous Carnegie RR Lyrae Program \citep[CRRP,][]{2012ApJ...758...24F} to leverage these excellent distance indicators and explore a variety of Local Group substructures including five dwarf galaxies (Sagittarius (Gupta et. al, in prep), Sculptor (Garofalo et. al, in prep), Ursa Minor, Carina, and Bo{\"o}tes) along with the Sagittarius and Orphan tidal streams. As we will show, the precision is such that we are able to resolve the three--dimensional structure of the stream, granting special access to a system that is in many ways the archetypal minor merger event.

In this work we present {\em Spitzer Space Telescope} \citep{2004ApJS..154....1W} Infrared Array Camera \citep[IRAC,][]{2004ApJS..154...10F} 3.6${\mathrm{ \mu m}}$ magnitudes and inferred distances to 32 candidate Orphan Stream RR Lyrae stars with the principal goal of informing future studies of the Galactic potential and Orphan progenitor. In Section~\ref{sec:obs} we describe our {\em Spitzer} photometry and the calculation of apparent magnitudes. Section~\ref{sec:distances} describes how we derive distances to individual Orphan stream stars. In Section~\ref{sec:fitting} we define a procedure to fit orbits to the RRL and measure bulk properties of the stream; in Section~\ref{sec:mass} we investigate the extent to which the orbit fits place constraints on the mass of the Milky Way. Section \ref{sec:prog} studies the Orphan progenitor and  Section~\ref{sec:concl} concludes.

\begin{figure*}
\includegraphics[width=\textwidth]{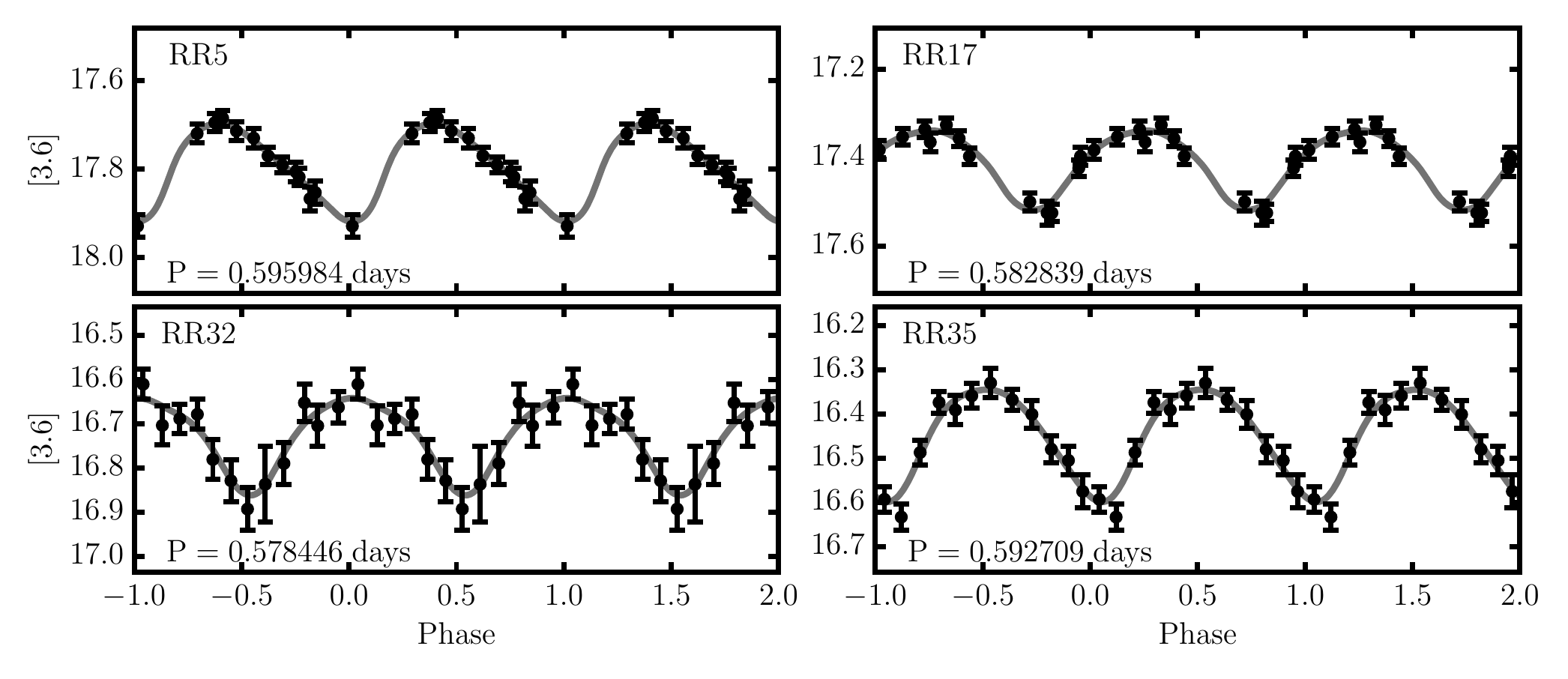}
\caption{Four example $\mathrm{3.6\mu m}$ SMHASH RRL lightcurves. The remainder are displayed in Figure~\ref{lcs}. The infrared lightcurves display a more sinusoidal shape than the sharply peaked and skewed optical lightcurves, as expected. This subset demonstrates the difference in phase coverage between the near and far subsamples; the more distant stars (RR5, RR17) may have substantial gaps resulting from the telescope's scheduling but often also smaller uncertainties in individual measurements due to the larger number of BCDs per epoch.\label{lc_sample}}
\end{figure*}

\section{Observations \& Data Reduction}
\label{sec:obs}
\subsection{Data Selection}
The RR Lyrae stars selected for observation in the SMHASH Orphan program are the 31 `high probability' candidate stream members of \cite{2013ApJ...776...26S}; these stars are all fundamental--mode pulsators (RRab). Also included is one `medium probability' candidate, RR5, because it was measured at large distance despite having a line--of--sight velocity somewhat discrepant with expectations for the Orphan stream given its position. The stars were identified from a compilation of three synoptic sky surveys: the Catalina Real--Time Sky Survey \citep[CRTS,][]{2009ApJ...696..870D}, the Lincoln Near Earth Asteroid Research \citep[LINEAR,][]{2000Icar..148...21S} survey, and the Palomar Transient Factory \citep[PTF,][]{2009PASP..121.1395L,2009PASP..121.1334R}. \cite{2013ApJ...776...26S} obtained follow--up spectroscopic observations in order to implement a Galactic standard of rest velocity cut as part of their stream membership criteria.  All of our targets therefore have uniformly determined metallicity \citep[on the Layden system, ][]{1994AJ....108.1016L} and line--of--sight velocity measurements with uncertainties of 0.15 dex and $\sim 15$ km/s, respectively. Their catalogue number in Table~\ref{datatable} is in order of decreasing declination, which approximately corresponds to a sequence of increasing apparent magnitude and decreasing Heliocentric distance (see Figure 2 in \citealt{2013ApJ...776...26S}.

\subsection{{\em Spitzer} Observations}
The mid infrared observations presented here were collected using the Infrared Array Camera (IRAC) on the {\textit Spitzer Space Telescope} as part of the Warm {\it Spitzer} Cycle 10  between 2014 June 19 and 2015 August 31 \citep{2013sptz.prop10015J}. Each star was observed in 12 epochs at 3.6~$\mu$m only. 

The targets selected in the Orphan stream span a wide range in distance, hence cover a significant range in apparent infrared magnitude. In order to achieve a sufficient signal--to--noise ratio on the individual epochs for both the nearest and most distant targets, the stars were divided into two groups based on their distances from \citet{2013ApJ...776...26S}, and their anticipated apparent magnitude from the K--band period--luminosity relation. The closer, brighter targets (with estimated distances less than $\sim 40$~kpc) were observed at each epoch with five dithered 100~s exposures, with all 12 epochs approximately uniformly spaced over a single pulsation cycle. The more distant, fainter targets used 25 dithered 100~s exposures to obtain the required S/N ratio. However, given the longer exposure times and the short pulsation cycle of the RRL, it was not possible to schedule all 12 observations within a single pulsation cycle. Instead these observations are spaced non--uniformly over several cycles, with typically 8--10 days between the first and last observation of a given target. 


\subsection{Photometry}
Individual Basic Calibrated Data (BCDs) generated by IRAC pipeline version S19.2 were downloaded from the {\em Spitzer} Science Center (SSC). Mosaics were created with the SSC--provided software {\sc mopex} \citep{2005ASPC..347...81M}; both individual-- and all--epoch (`master') mosaics for each field were produced with a 0.6" pixel scale. Point spread function (PSF) photometry was performed using the DAOPHOT/ALLSTAR/ALLFRAME program suite \citep{1987PASP...99..191S, 1994PASP..106..250S}. Further details of the SMHASH photometric procedure are provided in upcoming work (Garofalo et al., in prep). 

The Orphan Stream is highly diffuse so crowding from stream members is not important, but we find PSF photometry useful regardless to eliminate any contribution from field stars aligned by chance with the RRL. PSF stars are required to appear in at least 75\% of dithers and were chosen from uncrowded stars as determined by visual examination. For each target the PSF made from the epoch 1 mosaic is used on all epochs. Experiments with several stars showed no difference in measured magnitudes when using a PSF made from epoch 1, the master mosaic, or individual PSFs for each epoch.

The photometry was calibrated to the IRAC Vega magnitude system using the standard IRAC aperture correction procedure on the master mosaics, with inner and outer aperture radii of 6 and 14 pixels, respectively. Location corrections were applied to adjust for pixel--to--pixel sensitivity variations using the Warm Mission array location--correction images following the procedure outlined in the Warm {\it Spitzer} analysis documentation.

\subsection{Light Curves \& Average Magnitudes}

\label{sec:gloess}


The phase--folded lightcurves for each of our observed stars, using the period and time of maximum brightness determined from the optical data (Table~\ref{datatable}), are presented in Figure~\ref{lcs}; a subset is shown in Figure~\ref{lc_sample}. Each lightcurve is repeated for 3 phase cycles to highlight the variability. Stars where the telescope's scheduling resulted in multiple samples of the same point in phase (e.g. RR9, RR18) underscore {\em Spitzer's} precision photometric capabilities; field stars have a typical inter--epoch variation of approximately 0.03 mag, somewhat less than their single--epoch photometric uncertainty. Individual $3.6\mu$m magnitude measurement time series data for each star are provided as an electronic supplement to this article.

A smooth lightcurve is obtained from the observations using the Gaussian Local Estimation \citep[GLOESS,][]{2004AJ....128.2239P} algorithm. This technique evaluates the magnitude at a point in phase by fitting a second--order polynomial to the data, whose contributions to the fit are inversely weighted by the combination of both their statistical uncertainties and Gaussian distance from the point of interest. We use a Gaussian window of width 0.25 (in phase); the flux--averaged magnitude obtained from the fitted curve is not at all sensitive ($\Delta m = 1-3 \times 10^{-4}$ mag) to this smoothing length for any reasonable choice. The GLOESS lightcurve is used to determine the time--averaged, intensity--weighted mean magnitude. We compute the uncertainty on this quantity by adding in quadrature the per--star average photometric error and the uncertainty on the mean magnitude of the fitted lightcurve,

\begin{equation}
\sigma_{\mathrm{[3.6]}} = \sqrt{\frac{\Sigma\sigma_i^2}{N^2} +\sigma_{\rm fit}^2},
\end{equation}

\noindent where $N$ is the number of observations, $\sigma_{i}$ is an individual epoch's photometric uncertainty, and $\sigma_{\rm fit}$ is the uncertainty on the average magnitude calculated from the GLOESS fit. The latter is dependent on the observing scheme; one can show that the uncertainty on mean magnitude decreases as $1/N$ if the lightcurve is sampled uniformly, in contrast to the slower $1/\sqrt{N}$ drop for data that has been randomly sampled \citep{2012ApJ...758...24F}. Following the method of \cite{2011ApJ...743...76S}, we take advantage of this property where appropriate and compute $\sigma_{\rm fit} = A/(N\sqrt{12})$ for the brighter, uniformly sampled stars and $\sigma_{\rm fit} = A/(\sqrt{12N})$ for the fainter, nonuniformly observed subset, where A is the amplitude of the GLOESS lightcurve. Table \ref{datatable} compiles the SMHASH mean magnitudes calculated in this way along with the archival data.

\begin{table*}
\begin{threeparttable}
\caption{Distances and Light Curve Parameters of SMHASH Orphan RR Lyrae stars \label{datatable}}
\begin{tabular}{ccccccccccccccc}
ID & R.A. & Decl. & Period & $\mathrm{HJD_0}$ \tnote{a} & [3.6] mag \tnote{b}  &  [3.6] amp. & $\mathrm{A}_{[3.6]}$ \tnote{c}  & [Fe/H] & Helio. Distance \\
 & (J2000) & (J2000) & (days) & (days) & $\mathrm{mag}$ & $\mathrm{mag}$ & $\mathrm{mag}$ &  & $\mathrm{kpc}$\\
 \hline \\
RR4           & 142.596437 & 49.440867 & 0.677648 & 54265.667221  & 17.39 $\pm$ 0.01 & 0.158 & 0.003 & -2.32 & 44.04 $\pm$ 1.06 \\
RR5           & 139.486634 & 49.043981 & 0.595984 & 54508.734151  & 17.79 $\pm$ 0.02 & 0.223 & 0.002 & -2.05 & 48.88 $\pm$ 1.21 \\
RR6           & 143.840446 & 47.091109 & 0.530818 & 55887.972840  & 17.94 $\pm$ 0.03 & 0.341 & 0.002 & -2.37 & 50.91 $\pm$ 1.36 \\
RR7           & 141.771831 & 46.359489 & 0.639017 & 55590.054047  & 17.67 $\pm$ 0.02 & 0.257 & 0.003 & -1.94 & 47.27 $\pm$ 1.19 \\
RR9           & 144.271648 & 42.603354 & 0.567199 & 54913.653005  & 17.63 $\pm$ 0.02 & 0.219 & 0.002 & -2.08 & 44.36 $\pm$ 1.10 \\
RR10          & 142.541300 & 42.570500 & 0.649151 & 54157.679811  & 17.70 $\pm$ 0.02 & 0.211 & 0.002 & -2.53 & 50.62 $\pm$ 1.26 \\
RR11          & 144.881448 & 41.439236 & 0.624166 & 56271.888900  & 17.39 $\pm$ 0.02 & 0.200 & 0.002 & -2.56 & 43.26 $\pm$ 1.07 \\
RR12          & 146.057798 & 40.220714 & 0.711552 & 56334.821312  & 17.21 $\pm$ 0.02 & 0.228 & 0.003 & -2.35 & 41.61 $\pm$ 1.04 \\
RR13          & 143.482581 & 39.134007 & 0.527853 & 54415.904058  & 17.73 $\pm$ 0.02 & 0.186 & 0.002 & -2.22 & 45.47 $\pm$ 1.11 \\
RR14          & 143.913227 & 38.853250 & 0.504139 & 53789.793479  & 18.00 $\pm$ 0.01 & 0.151 & 0.002 & -2.36 & 51.21 $\pm$ 1.23 \\
RR15          & 146.447585 & 37.553258 & 0.624026 & 54913.654037  & 17.00 $\pm$ 0.02 & 0.183 & 0.002 & -2.14 & 34.87 $\pm$ 0.85 \\
RR16          & 148.586324 & 37.191956 & 0.573213 & 54941.722401  & 17.52 $\pm$ 0.01 & 0.151 & 0.002 & -2.18 & 42.81 $\pm$ 1.03 \\
RR17          & 142.909363 & 37.002696 & 0.582839 & 55598.766679  & 17.41 $\pm$ 0.02 & 0.179 & 0.002 & -2.73 & 43.01 $\pm$ 1.05 \\
RR18          & 146.008547 & 36.265846 & 0.594436 & 53789.812373  & 17.30 $\pm$ 0.01 & 0.163 & 0.002 & -2.27 & 39.53 $\pm$ 0.96 \\
RR19\tnote{d} & 146.390649 & 35.795310 & 0.755026 & 52722.727848  & 16.85 $\pm$ 0.01 & 0.051 & 0.002 & -1.96 & 34.92 $\pm$ 0.81 \\
RR23          & 150.579833 & 26.598017 & 0.573755 & 53078.770191  & 16.95 $\pm$ 0.03 & 0.313 & 0.004 & -2.42 & 33.61 $\pm$ 0.89 \\
RR24          & 150.243511 & 25.826153 & 0.708142 & 54476.844880  & 16.63 $\pm$ 0.01 & 0.158 & 0.005 & -2.14 & 31.17 $\pm$ 0.75 \\
RR25          & 150.647213 & 25.247547 & 0.542891 & 54539.656204  & 16.87 $\pm$ 0.02 & 0.202 & 0.005 & -2.12 & 30.83 $\pm$ 0.76 \\
RR26          & 151.892507 & 24.831492 & 0.620861 & 53788.855568  & 16.83 $\pm$ 0.02 & 0.231 & 0.006 & -2.09 & 32.09 $\pm$ 0.80 \\
RR27          & 150.544334 & 24.257983 & 0.604737 & 54595.657970  & 16.82 $\pm$ 0.02 & 0.267 & 0.005 & -1.86 & 30.89 $\pm$ 0.79 \\
RR29          & 153.996368 & 19.222735 & 0.645174 & 53816.785913  & 16.50 $\pm$ 0.02 & 0.234 & 0.004 & -2.00 & 27.84 $\pm$ 0.70 \\
RR30          & 153.698975 & 19.125864 & 0.630652 & 54149.788097  & 16.35 $\pm$ 0.02 & 0.175 & 0.005 & -2.09 & 25.86 $\pm$ 0.63 \\
RR31          & 154.238008 & 18.790623 & 0.508603 & 52648.880186  & 16.33 $\pm$ 0.02 & 0.235 & 0.005 & -1.97 & 23.06 $\pm$ 0.58 \\
RR32          & 154.824925 & 18.226018 & 0.578446 & 54084.925828  & 16.73 $\pm$ 0.02 & 0.220 & 0.005 & -1.61 & 28.42 $\pm$ 0.70 \\
RR33          & 154.469145 & 17.427796 & 0.575995 & 54207.717695  & 16.84 $\pm$ 0.02 & 0.219 & 0.005 & -1.75 & 30.16 $\pm$ 0.75 \\
RR34          & 154.295002 & 17.131504 & 0.513222 & 53706.970133  & 16.66 $\pm$ 0.02 & 0.253 & 0.005 & -1.88 & 26.71 $\pm$ 0.67 \\
RR35          & 156.791313 & 15.992450 & 0.592709 & 54175.771290  & 16.45 $\pm$ 0.02 & 0.254 & 0.005 & -2.32 & 26.84 $\pm$ 0.68 \\
RR39          & 158.493827 & 9.235715  & 0.554073 & 53851.699888  & 16.34 $\pm$ 0.02 & 0.219 & 0.004 & -2.00 & 24.13 $\pm$ 0.60 \\
RR43          & 160.996538 & 3.565153  & 0.618892 & 53710.968168  & 16.64 $\pm$ 0.02 & 0.231 & 0.006 & -2.31 & 29.87 $\pm$ 0.75 \\
RR46          & 161.045184 & 0.876656  & 0.591287 & 54535.792607  & 16.70 $\pm$ 0.03 & 0.295 & 0.007 & -1.58 & 28.26 $\pm$ 0.73 \\
RR47          & 161.622376 & 0.491299  & 0.463190 & 54180.766355  & 16.34 $\pm$ 0.02 & 0.263 & 0.006 & -1.50 & 21.31 $\pm$ 0.54 \\
RR49          & 162.349340 & -2.609458 & 0.523622 & 53054.827672  & 16.30 $\pm$ 0.02 & 0.245 & 0.006 & -2.02 & 23.05 $\pm$ 0.58 \\
\hline\\
\end{tabular}
\begin{tablenotes}
\item[a] Reduced Heliocentric Julian Date of maximum brightness (HJD -- 2400000)
\item[b] Extiction--corrected, flux--averaged  $\mathrm{3.6 \mu m}$ apparent magnitude from GLOESS fit (Section \ref{sec:gloess})
\item[c]  $\mathrm{3.6 \mu m}$ extinction from the \cite{2011ApJ...737..103S} dust map, calculated by \texttt{http://irsa.ipac.caltech.edu/applications/DUST/}
\item[d] RR19 is likely not an RR Lyrae star (or a member of the Orphan Stream) but we include it here for completeness.
\end{tablenotes}
\end{threeparttable}
\end{table*}

\subsection{Membership and contamination} \label{sec:RR19}
One of the principal difficulties in the study of halo substructure is separating tracers belonging to the object of interest from the background of halo objects of the same type. While the surveys contributing to the Orphan RRL catalogue are expected to be $> 95\%$ complete, partitioning the objects into members and contaminants is key to drawing any conclusions from them.  For this study of Orphan in particular, the issue is further complicated by one of Sagittarius' tails crossing the survey area around Galactic longitude $l \sim 200^\circ$; fortunately the Sagittarius debris is offset from the Orphan Stream in heliocentric radial velocity by $\sim 200$ km s$^{-1}$ in this part of the sky \citep[e.g.][]{2005ApJ...619..807L}. This section discusses several heuristics that may be used to differentiate individual populations.

A typical way of separating stellar systems is identifying characteristic patterns in their chemical abundances left by their star formation histories. Unfortunately, the SMHASH sample has a mean [Fe/H] of -2.1 dex and a dispersion of about 0.25 dex, which is not distinguishable from either the sample of stars in \cite{2013ApJ...776...26S} whose kinematics are inconsistent with stream membership or RR Lyrae stars more generally in the smooth halo \citep[mean $\mathrm{[Fe/H]} \sim -1.7, \sigma \sim 0.3$, ][]{2013ApJ...763...32D}. The mean metallicity can be used, however, to estimate how many Orphan stars we should expect in the survey area. Using the universal dwarf galaxy luminosity--metallicity relation obtained by \cite{2013ApJ...779..102K} and the Orphan Stream K--giant metallicity of $-1.63\pm0.19$ from \cite{2013ApJ...764...39C} (which should be more representative than the metal--poor RRL), we calculate that the progenitor should have had a luminosity $L_\mathrm{V} \sim 1.6 \times 10^6\ \mathrm{L_\odot}$. \cite{2016ApJ...818...41S} found that the quantity $\log_{10} N_{\mathrm{RRLy}}/\mathrm{L}_\odot$ is linear in metallicity with a scatter of 0.64 dex, which, when combined with the luminosity estimate, implies that the Orphan debris system has of order 100 RRL -- with an uncertainty of $\sim 0.7$ dex. Given that our precursor catalogues likely only cover one tail of the stream and that there are approximately 20 stars without spectra that \cite{2013ApJ...776...26S} find are consistent with the stream's distance, we conclude that the observed RRL population is appropriate given the probable progenitor.

Next we consider the contribution of a principal contaminant population -- the smooth stellar halo. For some time it has been known that the number density of halo RR Lyrae stars sharply decreases at a Galactocentric distance of approximately 25 kpc \citep{1985ApJ...289..310S}. More recent studies have shown that the power law index of this decline is $n = -4.5$ or greater \citep{2008ApJ...678..851K, 2009MNRAS.398.1757W, 2010ApJ...708..717S, 2017arXiv171001276C}. This is a significant advantage for studies of substructures beyond about 30 kpc as contaminants from the smooth component become almost negligible. For the case of the SMHASH Orphan footprint in particular, using the latest density normalization from \cite{2010ApJ...708..717S}, we expect only about 4 halo interlopers between 30 and 40 kpc and only 2 between 40 and 50 kpc; it is unlikely with such small numbers that they would also match the radial velocity trend of the stream. The catalogue star RR5 is marked as a medium--probability member for precisely this reason -- distant at 49~kpc but discrepant in radial velocity by 100 km s$^{-1}$.

There is also a subset of RRL that we do not expect to find as part of the Orphan Stream: high amplitude short period (HASP) RRab stars. These are fundamental mode pulsators that have large amplitudes, $A_\mathrm{V} \geq 0.75$ mag, but periods less than approximately 0.48 days. RR Lyrae variables in dwarf spheroidal galaxies do not populate this part of the period--amplitude plane, possibly because their metallicity evolution is too slow to produce a component both old enough and metal rich enough to pulsate in this range \citep{2002AJ....123..840B, 2015ApJ...798L..12F}. The smooth halo does, however, contain stars in the HASP parameter space at the several percent level and therefore such stars are likely contaminates. Amongst the SMHASH Orphan sample only RR47 meets the HASP criteria; it is also at the smallest distance from the Galactic centre, where the smooth halo is more dominant as described above. Since it has not yet been proven that the Orphan Stream's progenitor was a dwarf spheroidal galaxy we do not exclude RR47 from the following dynamical analyis but note that the conclusions are not substantively changed if it is omitted. 

Finally, we can use our $\mathrm{3.6 \mu m}$ data to identify non--RRL contaminants. Examination of the lightcurve for RR19 leads us to believe that it is not, in fact, an RRL. This star was observed over a single presumed period but there is no evidence of coherent variability. The optical lightcurve from the LINEAR, folded at the catalogue period, shows what might best be described as `bursty' variability, which is also inconsistent with being an RRL. Investigating this further, we performed our own period search on the LINEAR data and found no significant periods consistent with being an RRab for this star. We posit that this may simply be a false positive in the database. RR19 is therefore excluded from the rest of our analysis, however we include it in Table \ref{datatable} and Figure~\ref{lcs} for completeness.

\section{Distances to the Orphan RR Lyrae stars}
\label{sec:distances}

Distances to each Orphan RRL are determined using the (RRab--only) theoretical period--luminosity--metallicity (PLZ) relation of \cite{2017ApJ...841...84N}. They derived the PLZ using nonlinear, time--dependent convective hydrodynamical models of RR Lyrae variables with a range of metal abundances. They found that fitting those models with a simple period--luminosity relation results in an `intrinsic' scatter of $\sim 0.13$ mag, whereas including a metallicity term reduces the scatter to $\sim 0.035$ mag. The absolute magnitude in IRAC  $\mathrm{3.6 \mu m}$ is given by

\begin{equation}
\begin{split}
\mathrm{M}_{[3.6]} = {}& −2.276(\pm 0.021) \log(P) \\
&\quad +0.184 ( \pm 0.004) [\mathrm{Fe/H}] -0.786 ( \pm 0.007).
\end{split}
\end{equation}

\begin{figure}
\includegraphics[width=\linewidth]{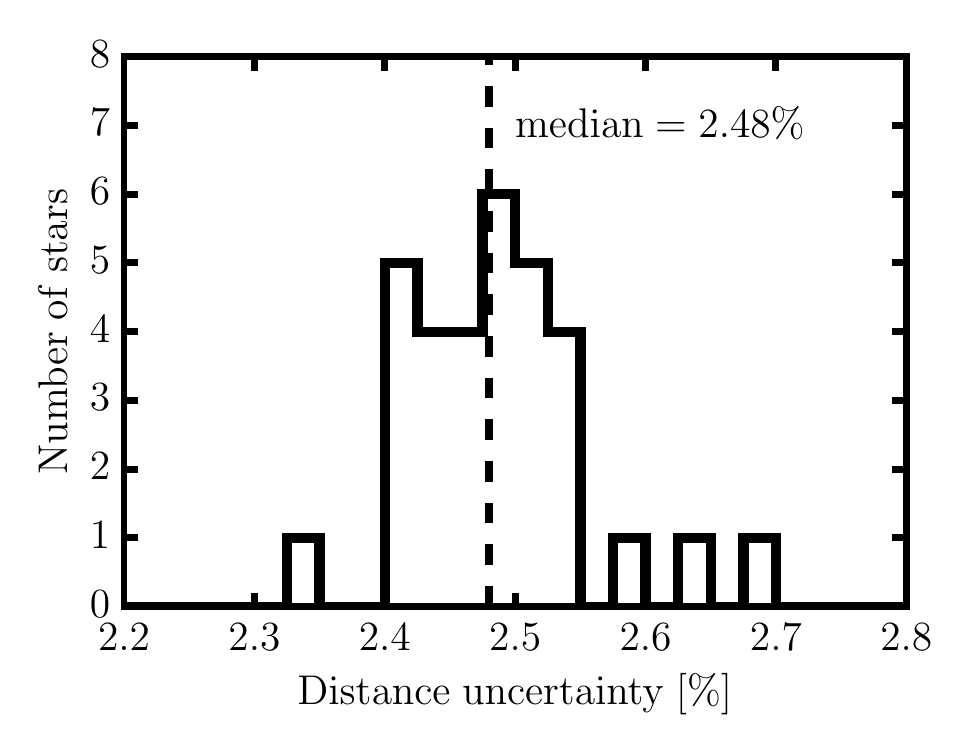}
\caption{\label{fig:unc} Relative heliocentric distance uncertainties $\sigma_d/d$ for the SMHASH Orphan RRL. The median uncertainty is indicated by the vertical dashed line. The scatter in the period--luminosity--metallicity relation and the uncertainty on apparent magnitudes each contribute $\sim1\%$, with the uncertainty in the star's individual metallicities providing the remainder.}
\end{figure}

\begin{figure}
\includegraphics[width=\linewidth]{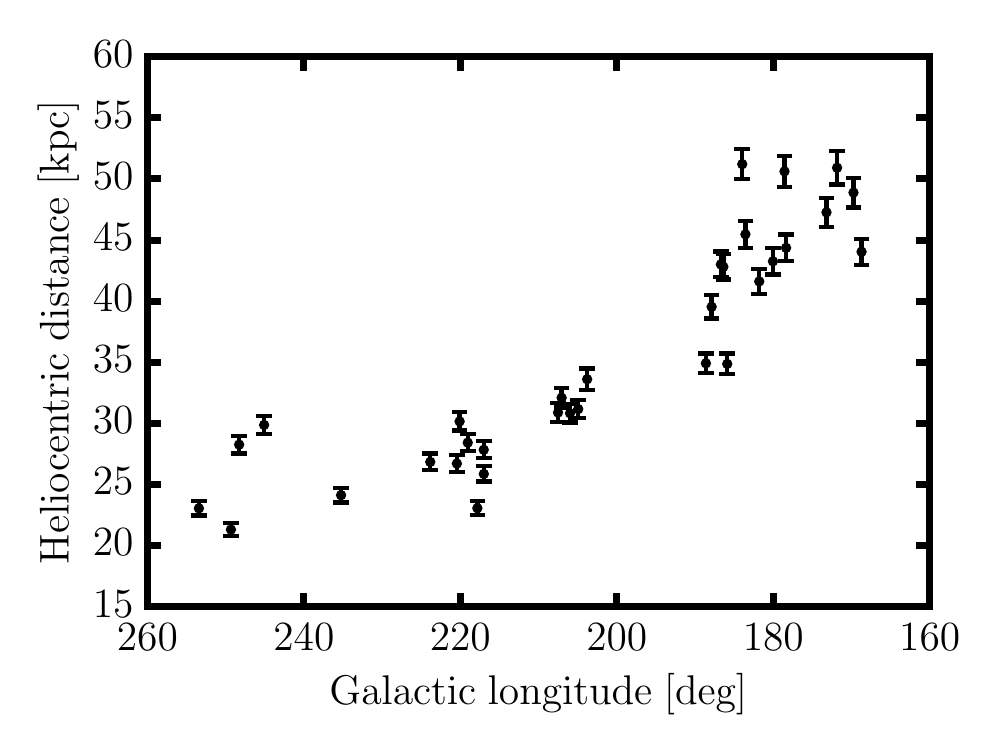}
\caption{\label{fig:dist} Measured heliocentric distances of the SMHASH Orphan RR Lyrae stars as a function of Galactic longitude. At large distances around 50 kpc the stream is approximately 8 kpc deep.}
\end{figure}

We fully propagate all sources of uncertainty, including those from the photometry, the lightcurve fit, the constants in the PLZ relation including its intrinsic scatter, the measured metallicities, and the extinction in this band, $A_{[3.6]}$. The latter is calculated from the \cite{2011ApJ...737..103S} dust map\footnote{evaluated using \\ \texttt{http://irsa.ipac.caltech.edu/applications/DUST/}}. Because the extinction is very low, $\sim 0.005$ mag, the entire value is adopted as the uncertainty on extinction. This conservative choice negligibly affects the resultant uncertainty on $M_{[3.6]}$. 

The SMHASH Orphan Stream sample's distance uncertainty distribution is shown in Figure~\ref{fig:unc}. The median relative distance uncertainty is a mere 2.5\% For comparison, end--mission parallax distances to RR Lyrae stars obtained by {\it Gaia} are expected to have 10\% uncertainties for stars at just 6 kpc \citep{2013ApJ...778L..12P}, while we are measuring stars at 51 kpc. 

It is interesting to consider which, if any, of the observational uncertainties most strongly limit the precision of SMHASH distances. An elementary analysis of the error budget suggest that the metallicity uncertainty and Z term slope contribute 0.5\%, the photometric and fit uncertainties contribute 0.9\%, and the intrinsic scatter, period slope and zero point are responsible for 1.1\% of the 2.5\% relative uncertainty. The heliocentric distances derived for each RRL using the \cite{2017ApJ...841...84N} PLZ relation are given in Table~\ref{datatable}.

Figure~\ref{fig:dist} shows the RRLs' heliocentric distances as a function of Galactic longitude. We trace the stream to approximately 51 kpc. This figure makes it apparent that the Orphan Stream is not `thin' at large distances; near $l=180^\circ$ the stream is approximately 10 kpc deep from a heliocentric perspective. In Section~\ref{sec:prog} we will argue that this depth contains information about the stream's progenitor. Overall, the SMHASH distances are in good agreement ($\sim 1\sigma$) with the previous work of \cite{2013ApJ...776...26S}, who used an optical luminosity--metallicity relation \citep{2003LNP...635..105C} to obtain distances to these same RR Lyrae stars. On average we find that our measurements are 5\% larger than the values of \cite{2013ApJ...776...26S}; notably, however, we find that their two most remote stars are $\sim 5$ kpc closer, reducing the maximum heliocentric distance of the stream from about 55 to about 51 kpc.

\section{Properties of the Orphan Stream}
\label{sec:fitting}

In the following we assume that all of the SMHASH RR Lyrae stars do indeed belong to the Orphan Stream, and so use them to outline its path and properties. We do this by (i) assuming a form for a galactic potential; (ii) finding the parameters of the potential and the orbit within that potential that best fits the centroid of the RRL positions in their measured dimensions; and (iii) measuring the dispersions in line--of--sight distance, angular size on the sky, and radial velocity about this best--fitting orbit.

Note that, since orbits of debris stars are offset from the progenitor satellite orbit \citep{1998ApJ...495..297J,1999MNRAS.307..495H}  we expect this approach to provide {\it biased} estimates of the true potential parameters and orbit of the progenitor  \citep[see][as well as our own exploration in Section \ref{sec:bias}]{2011MNRAS.413.1852E, 2013MNRAS.433.1813S, 2013MNRAS.436.2386L}. We nevertheless choose to fit orbits and potentials rather than -- for example -- a polynomial to the path since this allows us to both measure the structure of the stream via its depth and compare our results to the prior work of \cite{2010ApJ...711...32N}. The reader is cautioned that the `best--fitting' potential and orbit are not expected to correspond exactly to the potential of the Milky Way or the orbit of the progenitor. However, the dispersion about the path outlined by the stream {\it do} contain clues to the nature of the progenitor (see Section~\ref{sec:prog}).

\subsection{Fitting method}

To fit an orbit to our RRL we use \texttt{emcee} \citep{2013PASP..125..306F}, a Python implementation of an affine--invariant ensemble sampler for a Markov Chain Monte Carlo (MCMC) algorithm \citep{GW2010}, to draw samples from the posterior probability density of the model parameters. This method is similar to that of \cite{2010ApJ...712..260K}, \cite{2015ApJ...809...59S} and \cite{2016ApJ...824..104P}.

\subsubsection{Potential model}
\label{sec:pot}

The Milky Way potential is represented as three smooth, static components: a \cite{1975PASJ...27..533M} disk, a Hernquist \citep{1990ApJ...356..359H} bulge, and a spherical logarithmic halo, defined as

\begin{equation}
\label{eq:disk}
\Phi_{\rm disk} = - \alpha \frac{GM_{\rm disk}}{\sqrt{R^2 + (a + \sqrt{z^2+b^2})^2}}
\end{equation}
\begin{equation}
\Phi_{\rm sphere} = \frac{GM_{\rm sphere}}{r+c}
\end{equation}
\begin{equation}
\label{eq:halo}
\Phi_{\rm halo} = v_{\rm halo}^2 \ln{\left( R^2 + z^2 + r_h^2 \right)}
\end{equation}

\noindent with component masses $M_{\rm sphere} = 3.4 \times 10^{10}\mathrm{M_\odot}$ and $M_{\rm disk} = 1 \times 10^{11} \mathrm{M_\odot}$, disk scale length $a = 6.5$ kpc, disk scale height $b = 0.26$ kpc, bulge core radius $c = 0.7$ kpc, and halo scale radius $r_h = 12$ kpc; $R$ and $z$ are the cylindrical coordinates and $r$ is the spherical radius. We fix the solar distance to the Galactic centre as $R_0 = 8$ kpc \citep[consistent with previous work, but also measurements e.g.][]{2009ApJ...707L.114G} and the peculiar velocity of the Sun (U,V,W)$_\odot =  (11.1,\ 12.24,\ 7.25)$ km s$^{-1}$ \citep{2010MNRAS.403.1829S}. In the orbit fitting algorithm the only potential parameter allowed to vary is the dark matter halo's scale velocity $v_{\rm halo}$, with $r_h$ chosen such that the total potential's circular velocity at the solar position is $220\ \mathrm{km\ s^{-1}}$ \citep[e.g.][]{2012ApJ...759..131B}. These parameters are chosen to match Model 5 of \cite{2010ApJ...711...32N} (their best--fitting model with a logarithmic halo) which in turn is an implementation of the best--fitting spherical model of \cite{2005ApJ...619..807L} except that the halo scale velocity is allowed to vary. We note that the constraint on the circular velocity precludes us from fitting precisely \citet{2010ApJ...711...32N}'s Model 5 since that potential's circular velocity at the solar position is only $207$ km s$^{-1}$.

\subsection{Model parameters}

We wish to find the phase space coordinates of the initial condition ${\bf x^0} = (l,b,DM,\mu_l, \mu_b, v_r)^0$ for the orbit that best reproduces the observed sky positions $l_i, b_i$, heliocentric radial velocities $v_{r,i}$ and distance modulii $DM_i$ of the RRL given their uncertainties $\sigma_{v_{r,i}}, \sigma_{DM_{i}}$. The sky coordinates are assumed perfectly known and are transformed to the Orphan frame $\Lambda, B$ defined in \cite{2010ApJ...711...32N}, a heliocentric spherical coordinate system in which the Orphan Stream lies approximately on the equator. The rotation between Galactic coordinates and the Orphan coordinates is defined by the Euler angles $(\phi, \theta,\psi)= (128.79^\circ,54.39^\circ,90.70^\circ)$. We set $l^0 = 200^\circ$ without interesting loss of generality.

Because tidal streams are generated with orbital parameters somewhat offset from the progenitor galaxy and with some intrinsic scatter \citep[cf.][and references therein]{2015MNRAS.454.2472H} we also include additional model parameters ${\bf \delta} = (\delta_{B}, \delta_{v_{r}}, \delta_{DM})$ to account for the average dispersions in the observational coordinates. We neglect the fact that each of these dispersions will vary along the stream. Besides representing the physical width, velocity dispersion, and depth of the stream, they serve to deter over--fitting in coordinates where $\delta/\sigma$ is large. The last parameter is the halo scale velocity $v_{\rm halo}$. The full parameter set is then ${\bf\theta} = ((b,DM,\mu_l, \mu_b, v_r)^0, (\delta_{B}, \delta_{v_{r}}, \delta_{DM}),v_{\rm halo})$. Orbits were integrated using a symplectic leapfrog integrator as implemented in the Gala package \citep{gala}

The MCMC algorithm uses 144 walkers to explore this nine--dimensional parameter space. After running for a burn-in period of 1,000 steps the sampler is restarted and run for an additional 10,000 steps. Since the autocorrelation time for each walker is $\sim$~50 steps in all dimensions, only every 100th sample is taken from the chains to be included in the posterior. This ensures that each is a nearly independent sample from the posterior distribution. The autocorrelation time does not change substantially after the burn-in period, indicating that the sampling has converged.

\subsubsection{Likelihood}

\begin{figure*}
\includegraphics[width=\linewidth]{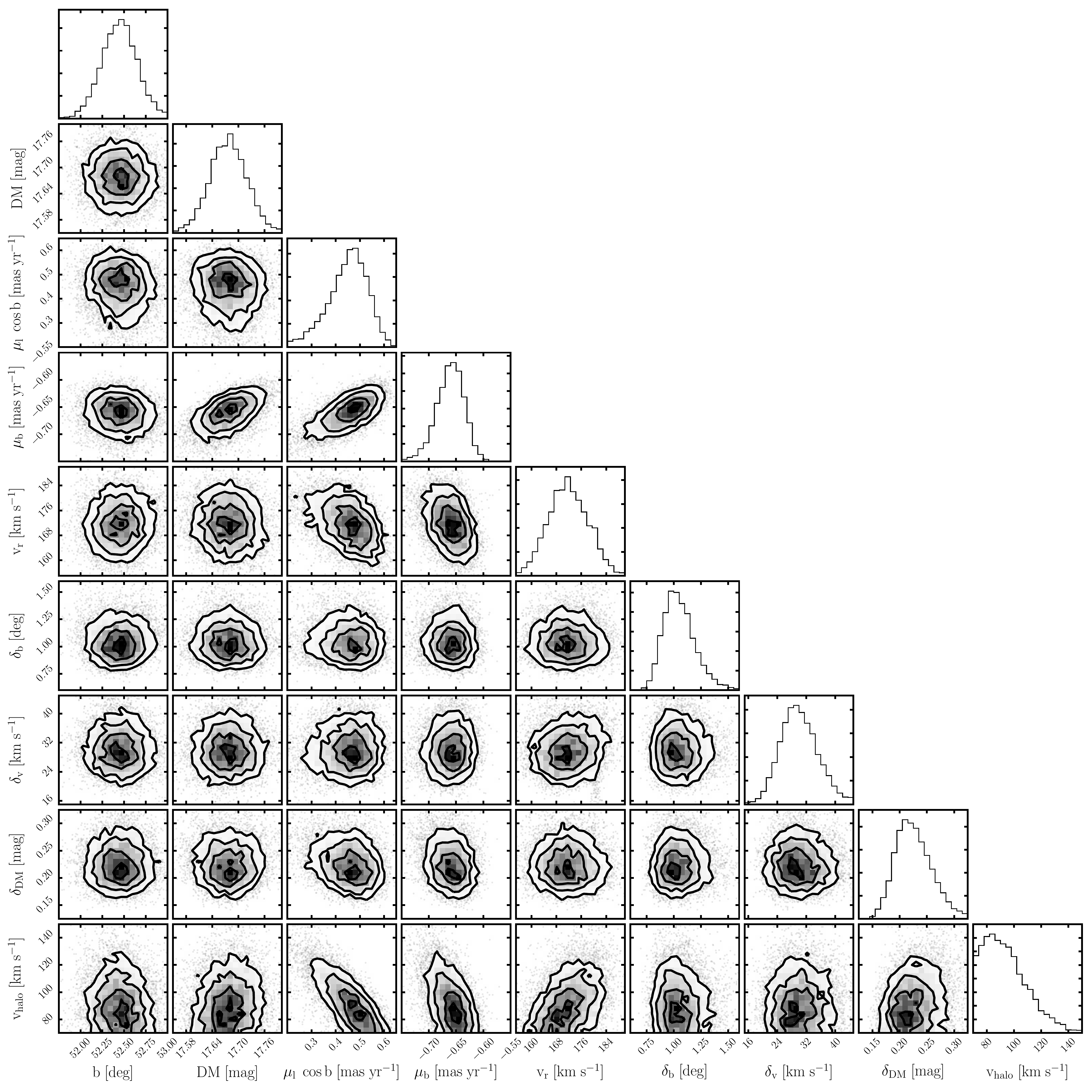}
\caption{\label{fig:corner} Corner plot \citep{corner} displaying the posterior distributions of the MCMC walkers for the case of uninformative proper motion priors. Contour plots show the posterior marginalized over the other seven dimensions; histograms are marginalized over all but one. In general there is little covariance between parameters with the notable exceptions of $\mu_l\cos{b}$ with $\mu_b$ and the velocity components with $v_{\rm halo}$. This suggests that precise proper motion measurements will add significantly to the constraint on enclosed mass.}
\end{figure*}

We assume that our data are independent and that the uncertainties in each coordinate are normally distributed. Thus the joint likelihood is the product of the likelihoods in each coordinate, which are
\begin{equation}
p(B_i | \Lambda_i, {\bf{\theta}}) = \mathcal{N}(B_i | B^{\rm model}(\Lambda_i), \delta_B^2)
\end{equation}
\begin{equation}
p(v_{r_i} | \Lambda_i, {\bf{\theta}}) = \mathcal{N}(v_{r_i} | v_r^{\rm model}(\Lambda_i), \sigma^2_{v_r} + \delta^2_{v_r})
\end{equation}
\begin{equation}
p(DM_i | \Lambda_i, {\bf{\theta}}) = \mathcal{N}(DM_i | DM^{\rm model}(\Lambda_i), \sigma^2_{DM} + \delta^2_{DM})
\end{equation}
where $B^{\rm model},~v_r^{\rm model}$, and $DM^{\rm model}$ are interpolated from the model orbit integrated using the initial conditions in $\theta$ and $\mathcal{N}$ is the normal distribution 
\begin{equation}
\mathcal{N}(x | \mu, \sigma^2) = \frac{1}{\sqrt{2\pi\sigma^2}}\exp{-\frac{(x-\mu)^2}{2\sigma^2}}
\end{equation}

\noindent with $\mu$ as its mean and $\sigma$ its standard deviation.

\subsubsection{Priors}

We implement priors on Galactic latitude and distance modulus that are uniform in Cartesian space; for the former this is uniform in $\cos(b)$, while the latter is 
\begin{equation}
p(\mathrm{DM}) \propto 10^{\frac{2}{5}\mathrm{DM}+2}.
\end{equation}
Using the notation $\mathcal{U}(f,g)$ for the uniform distribution with endpoints $f$ and $g$, we place an uninformative prior on Heliocentric radial velocity as 
\begin{equation}
p(v_r) = \mathcal{U}(50,300)\ \mathrm{km\ s^{-1}}.
\end{equation}
\noindent The dispersions $\delta_i$ are required to be positive to prevent a physically equivalent but bimodal posterior that hampers the walkers' convergence. We use logarithmic (scale-invariant) priors for these parameters,
\begin{equation}
p(\delta_{i}) \propto \delta_{i}^{-1}
\end{equation}
\noindent The halo scale velocity $v_{\rm halo}$ must be greater than about 68 km s$^{-1}$ to maintain a circular speed at the solar radius of 220 km s$^{-1}$ given our choices for the other parameters. It is therefore constrained by
\begin{equation}
p(v_{\mathrm{halo}}) = \mathcal{U}(68,200)\ \mathrm{km\ s^{-1}}.
\end{equation}
\noindent Finally, we consider the two phase space dimensions that are unobserved for individual RRL: their proper motions. Since we cannot compare them to a prior on a star--by--star basis, we instead use the value for the model orbit where it crosses $l=199.7796^\circ$. This position is specifically chosen to correspond to the location of {\it Hubble Space Telescope} -- based proper motions of Orphan Stream stars \citep{2016ApJ...833..235S}. We consider two cases: first wide, uninformative priors
\begin{equation}
p(\mu_l\ \cos{b}) = \mathcal{U}(-5,5)\ \mathrm{mas\ yr^{-1}}
\end{equation}
\begin{equation}
p(\mu_b) = \mathcal{U}(-5,5)\ \mathrm{mas\ yr^{-1}},
\end{equation}
\noindent and then those based on the {\it Hubble} observations
\begin{equation}
p(\mu_l\ \cos{b}) = \mathcal{N}(0.211,0.05^2)\ \mathrm{mas\ yr^{-1}}
\end{equation}
\begin{equation}
p(\mu_b) = \mathcal{N}(-0.774,0.05^2)\ \mathrm{mas\ yr^{-1}}.
\end{equation}

\noindent In the following we will refer to the former as `without' a proper motion prior for conciseness.

\begin{figure*}	
	\centering
	\begin{subfigure}[t]{.47\linewidth}
		\centering
		\includegraphics[width=\linewidth]{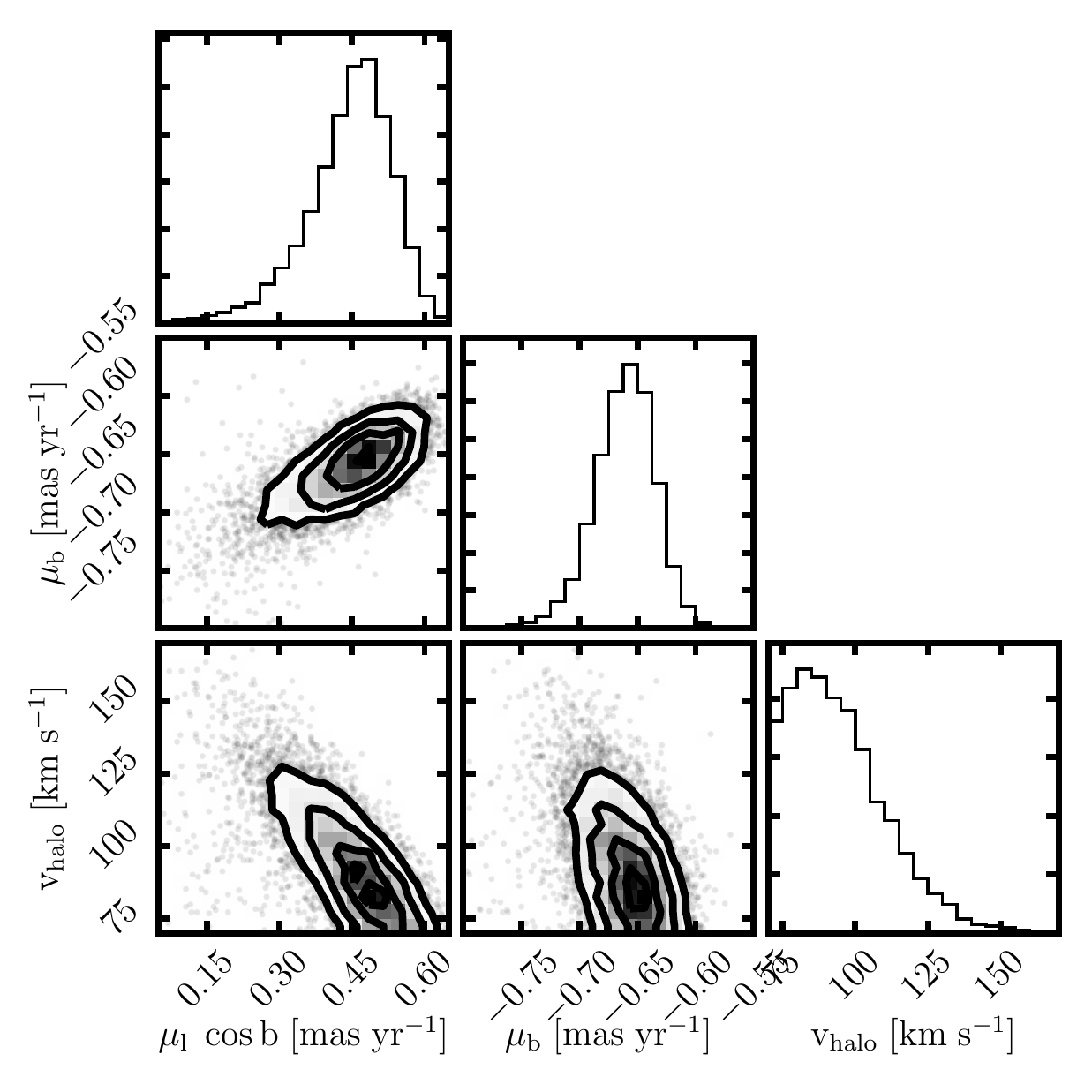}
		\label{fig:corner_comp_nopm}		
	\end{subfigure}
	\quad
	\begin{subfigure}[t]{.47\linewidth}
		\centering
		\includegraphics[width=\linewidth]{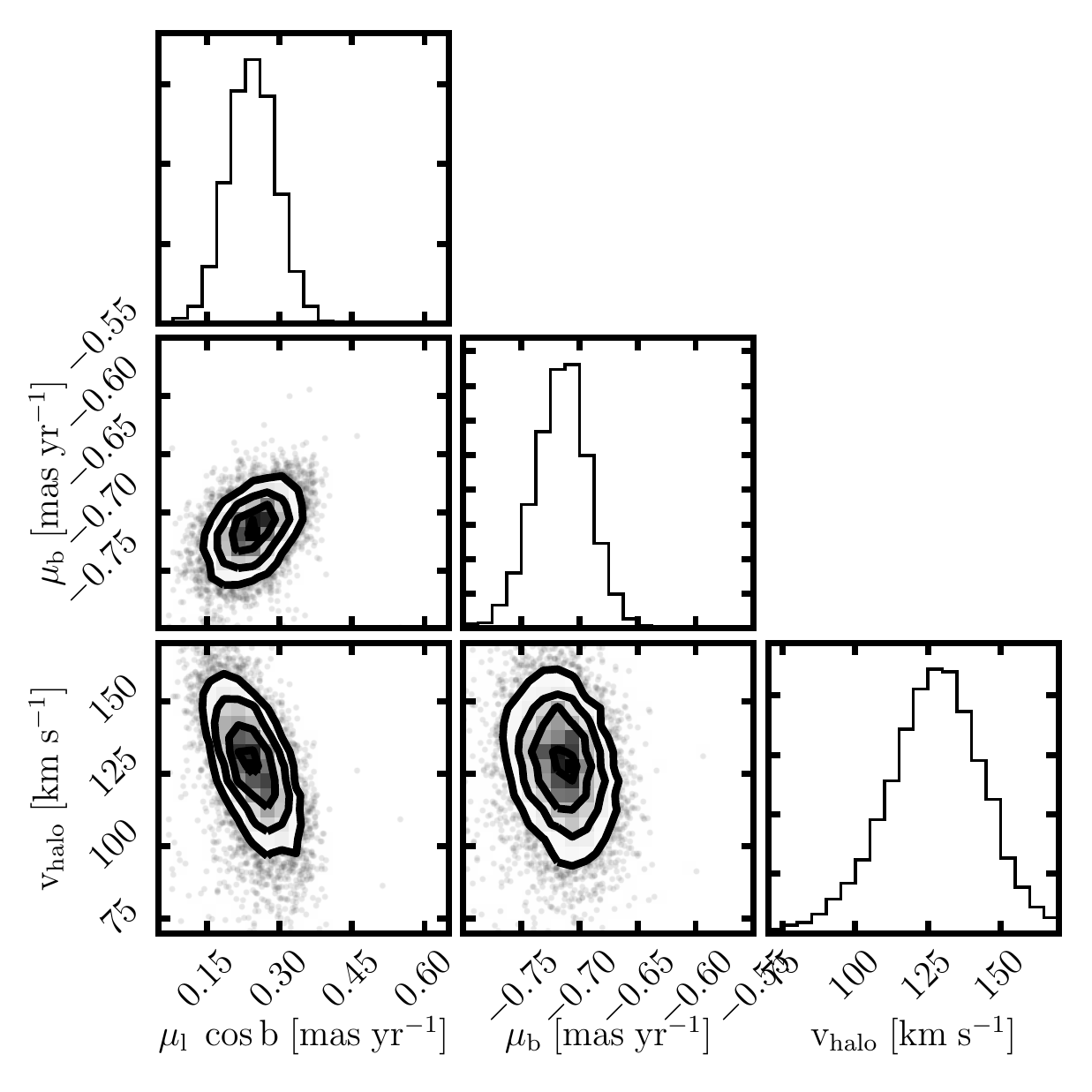}
		\label{fig:corner_comp_pm}
	\end{subfigure}
	\caption{Corner plot displaying the marginalized posterior distributions for the model parameters $\mu_l\cos(b)$, $\mu_b$, and $v_{\rm halo}$ along with their covariances. Left: uniform prior on $\mu_l\cos(b)$ and $\mu_b$. Right: result when otherwise identical chains are run with the additional priors $p(\mu_l \cos(b))=\mathcal{N}(0.211,0.05^2)$, $p(\mu_{b})=\mathcal{N}(-0.774,0.05^2)$. Due to the covariance between the proper motions and the halo scale velocity, these priors result in a median $v_{\rm halo}$ that corresponds to a halo 64\% more massive than the uniform case.}
	\label{fig:corner_pm_comparison}
\end{figure*}

\subsection{Centroid of the Orphan Stream}

Figure~\ref{fig:corner} shows a corner plot displaying projections of the orbit fitting's posterior distribution, in the case of the uniform proper motion priors. The median value of the samples in each parameter, along with uncertainties computed as the 16th and 84th percentiles (the 68\% credible interval), are summarized in Table~\ref{tab:fitpars}. We confirm that the orbit is prograde with respect to the Milky Way's rotation. Even if the walkers are restricted to only exploring the space of retrograde orbits, there are no local maxima to compare to the prograde fit shown here. If the overdensity detected by \cite{2015ApJ...812L..26G} is indeed the nearly--disrupted progenitor then this direction of motion makes the SMHASH RR Lyrae stars part of the leading tidal tail. The median distance modulus of 17.68 mag corresponds to a heliocentric distance of 34.2 kpc; this is approximately 150 pc more distant than \cite{2010ApJ...711...32N}'s Model 5 orbit at the same longitude, however they are compatible within their respective uncertainties.

Focusing on each of the 2d histograms in Figure~\ref{fig:corner} in turn, one sees that the fit parameters have minimal covariance with few exceptions: the proper motions $\mu_l\cos(b)$ with $\mu_b$, $v_{\rm halo}$ with $\mu_l\cos(b)$, and to a lesser extent $v_{\rm halo}$ with $\mu_b$ and with $v_r$. Note that the stream's Galactic latitude varies by only a few degrees in the area of our observations. It is no coincidence that the velocity components covary with the scale of the halo; it represents the need for additional kinetic energy to reach the same Galactocentric radius in a deeper potential. This means that currently available proper motion measurements can be highly informative when applied in combination with SMHASH's precision distances. For example, the 68\% credible interval of the marginalized posterior for $\mu_l\cos(b)$ spans almost 0.2 mas yr$^{-1}$ while the uncertainty on the same quantity computed from the measurement of \cite{2016ApJ...833..235S} is $\approx$ 0.05 mas yr$^{-1}$.

\subsection{Stream fitting with six-dimensional constraints}

Figure~\ref{fig:corner_pm_comparison} illustrates the effect of the precise proper motion constraints on the final positions of the MCMC walkers on the three most affected dimensions -- $\mu_l\cos(b)$, $\mu_b$, and $v_{\rm halo}$. On the left we highlight these quantities in the uninformative case; here we find $\mu_l\cos(b)$ and $\mu_b$ from the best--fitting orbits are $\sim 2\sigma$ discrepant with the measured value. The strength of the \cite{2016ApJ...833..235S} priors are such that when applied to the walkers (on the right) the means of the marginalized posterior distributions are shifted wholesale, making the two nearly disjoint. The halo scale parameter is dragged to significantly higher values, as one would naively expect based on the covariance with $\mu_l\cos(b)$. 

\begin{table}
\renewcommand{\arraystretch}{1.5}
\begin{tabular}{lcc}
Parameter & Without PM prior & With PM prior\\
\hline
$\mathrm{l\ [deg]}$  & 199.7796 & 199.7796 \\
$\mathrm{b\ [deg]}$  & $\mathrm{52.45_{-0.21}^{+0.21}}$ & $\mathrm{52.46_{-0.21}^{+0.23}}$ \\
$\mathrm{DM\ [mag]}$  & $\mathrm{17.68_{-0.04}^{+0.04}}$ & $\mathrm{17.66_{-0.05}^{+0.05}}$ \\
\\
$\mathrm{\mu_l\ \cos(b)\ [mas\ yr^{-1}]}$ & $\mathrm{0.456_{-0.096}^{+0.071}}$ & $\mathrm{0.244_{-0.051}^{+0.049}}$ \\
$\mathrm{\mu_b\ [mas\ yr^{-1}]}$  & $\mathrm{-0.660_{-0.028}^{+0.023}}$ & $\mathrm{-0.715_{-0.024}^{+0.022}}$ \\
$\mathrm{v_r\ [km\ s^{-1}]}$ & $\mathrm{171.7_{-6.3}^{+6.9}}$ & $\mathrm{176.2_{-6.8}^{+6.5}}$ \\
\\
$\mathrm{\delta_B\ [deg]}$  & $\mathrm{1.042_{-0.129}^{+0.168}}$ & $\mathrm{1.039_{-0.129}^{+0.175}}$ \\
$\mathrm{\delta_v\ [km\ s^{-1}]}$ & $\mathrm{29.86_{-4.82}^{+5.72}}$ & $\mathrm{29.61_{-4.94}^{+5.81}}$ \\
$\mathrm{\delta_{DM}\ [mag]}$  & $\mathrm{0.224_{-0.030}^{+0.040}}$ & $\mathrm{0.258_{-0.036}^{+0.046}}$ \\
\\
$\mathrm{v_{halo}\ [km\ s^{-1}]}$ & $\mathrm{92_{-14}^{+19}}$ & $\mathrm{128_{-17}^{+16}}$ \\
$\mathrm{M(60\ kpc)\ [10^{11} M_\odot]}$ & $\mathrm{3.4_{-0.65}^{+1.1}}$ & $\mathrm{5.6_{-1.1}^{+1.2}}$ \\
\hline
\end{tabular}
\caption{Median and 68\% credible intervals of parameters in the posterior distribution resulting from orbit fitting to the SMHASH data, with and without including the observational proper motion constraints. The fixed Galactic longitude value used for the initial condition is included for completeness, along with the mass enclosed at 60 kpc implied by the $v_{\rm halo}$ distribution.\label{tab:fitpars}}
\end{table}

\begin{figure}
\includegraphics[width=\linewidth]{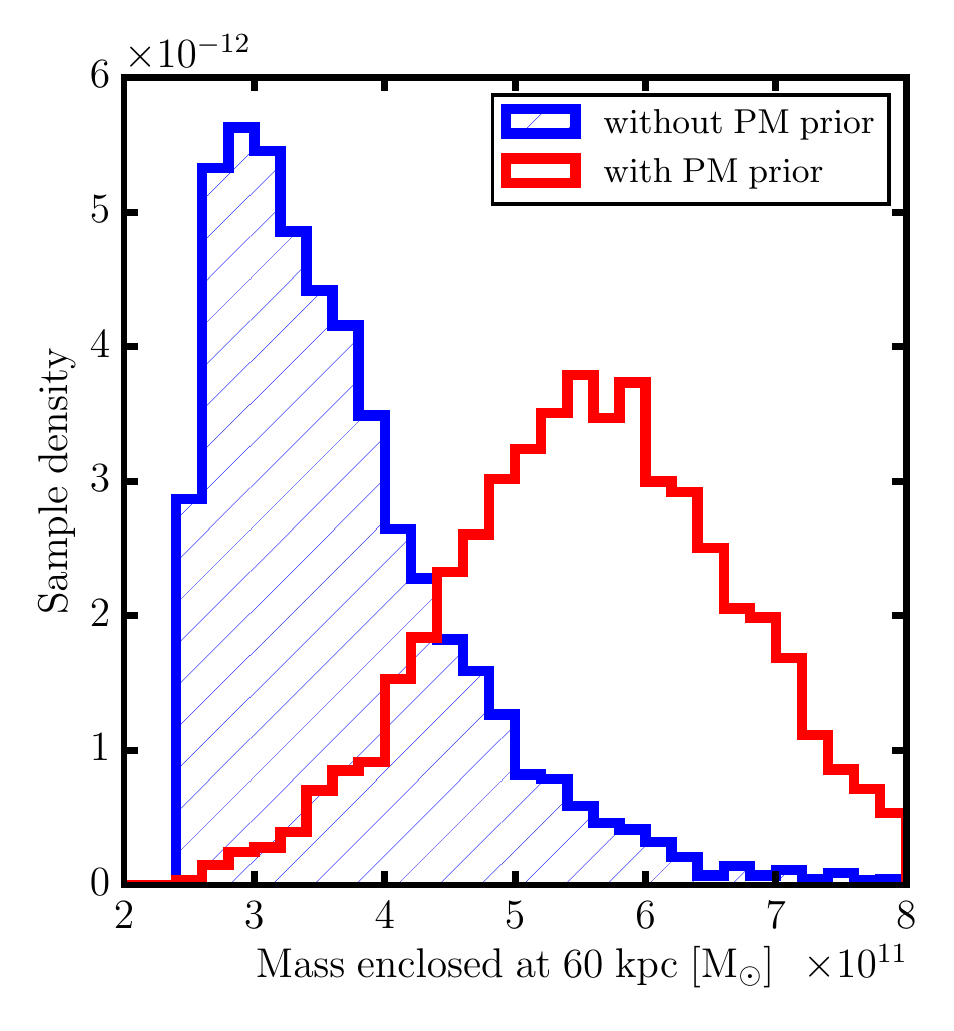}
\caption{\label{fig:masshist} Milky Way mass enclosed at 60 kpc, calculated from the scale velocities $v_{\rm halo}$ of the samples. Including the proper motion prior significantly increases the median value, from $\mathrm{3.4\times 10^{11} M_\odot}$ (in blue hatch) to $\mathrm{5.6 \times 10^{11} M_\odot}$ (in red)}. 
\end{figure}

The marginalized posterior of $v_{\rm halo}$ can be directly converted into a distribution of enclosed masses at any given radius; we choose 60 kpc for convenient comparison with literature values. The results of this transformation are shown in Figure~\ref{fig:masshist}, both without (in blue hatch) and with (in red) the observed proper motions as a prior. The difference between them is dramatic: the latter's median value is 64 per cent larger than the former.

\begin{figure*}
\includegraphics[width=\linewidth]{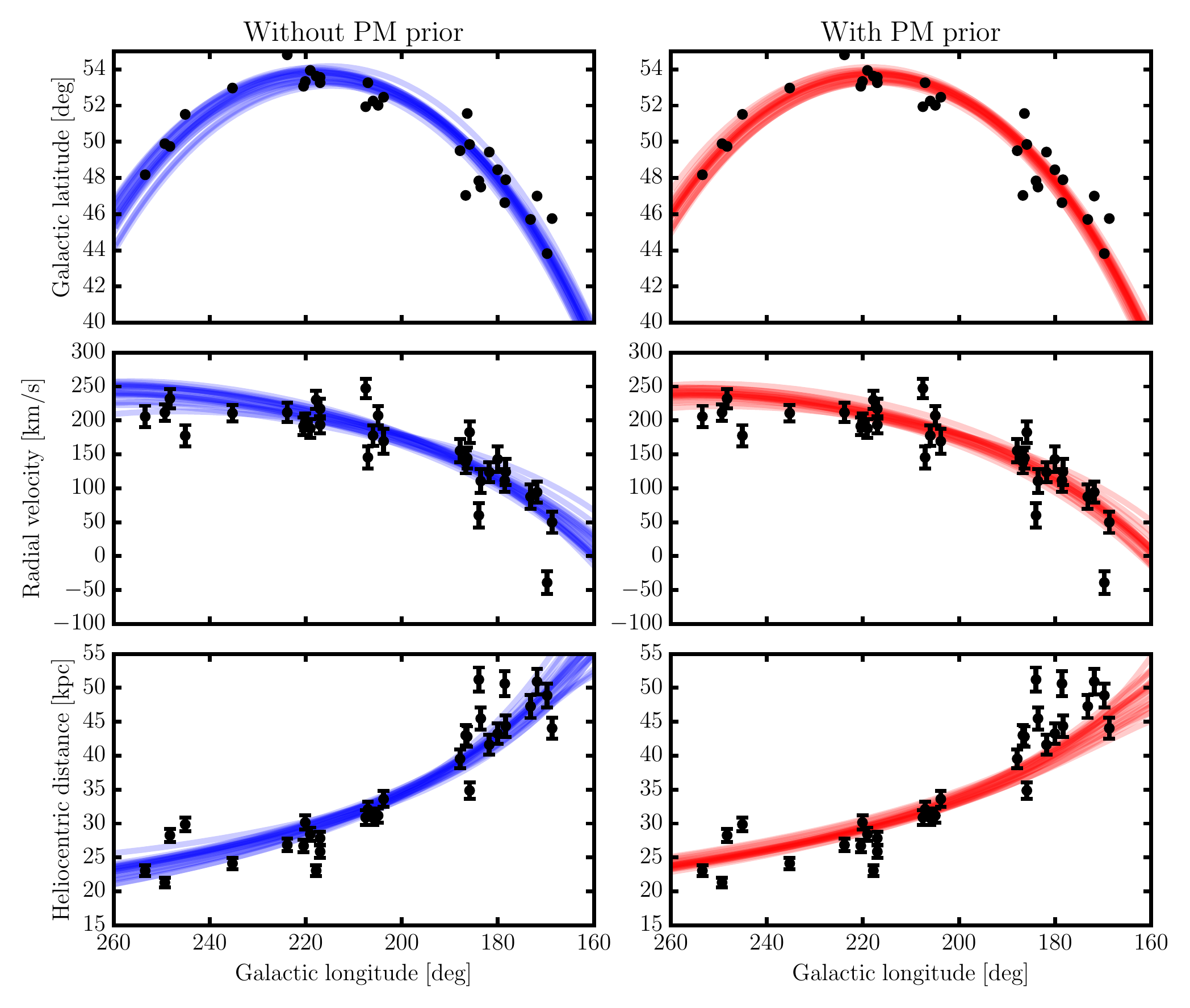}
\caption{\label{fig:orbits_jill} Left: a selection of orbit fits (blue lines) generated from randomly selected samples of the posterior distribution shown in Figure~\ref{fig:corner}, where the proper motion prior is uninformative. Right: the same (red lines), but with samples from the walkers constrained by the observed $\mu_l\cos{b}$ and $\mu_b$. The former better reproduces the trend of distance with longitude, while the latter slightly improves the match in radial velocity and sky position, especially at $l > 240^\circ$.}
\end{figure*}

A selection of orbits generated from randomly chosen samples of the posteriors are shown in Figure~\ref{fig:orbits_jill}. The left (right) panels show the results without (with) including informative proper motion priors. Plotted from top to bottom are projections in the three observational coordinates (Galactic latitude, radial velocity, and distance) as a function of Galactic longitude. Both sets of samples capture the path of the stream over most of the survey area.  Individual orbits diverge somewhat around $l \lesssim 170^{\circ}$ where the depth in line--of--sight distance is large. Both sets of orbits seem to systematically overestimate the Heliocentric radial velocity of stars above $l \approx 250^\circ$, however individual stars are only offset by $\sim 1\ \delta_i$. Including the \cite{2016ApJ...833..235S} measurement slightly improves the match to the data in $b$ and $v_r$ but causes the distance to the far end of the stream to be underestimated. This is problematic because the leading arm of the stream is made up of stars with lower specific energy than the progenitor and are expected to be {\it interior} to its orbit. We interpret this mismatch as evidence that the 1--parameter potential model used here is not flexible enough to recover the full phase space structure of the stream. In the N--body models described below there is no offset between fitted orbits and selected particles at the 0.05 mas yr$^{-1}$ level. 

\subsection{The Solar circular velocity as measured from the Orphan Stream}

To the extent that a stream follows an orbit, the proper motion of member stars perpendicular to the stream should be zero. Any observed perpendicular proper motion is therefore a measure of the solar reflex \citep[c.f.][]{2012ApJ...744...25C}. The {\it Hubble} proper motion measurement and the SMHASH distance distribution posterior can be combined at the longitude of the \cite{2016ApJ...833..235S} Orphan F1 field to estimate the solar motion.  

We define a new coordinates system relative to the Orphan coordinates of \cite{2010ApJ...711...32N} with axes that point into the plane of the sky, parallel to the stream, and perpendicular to the stream. The unit vector perpendicular to the stream points in the direction (in Orphan coordinates) $\hat{n} = (0.62619,\  0.50664,\  0.59261)$. In this direction, the marginalized posterior derived using the {\it Hubble} proper motion priors approximates a Gaussian with mean 136.5 km s$^{-1}$ and dispersion 9.1 km s$^{-1}$. If we assume that the solar peculiar velocity relative to the local standard of rest (LSR) is known from \cite{2010MNRAS.403.1829S}, then this implies that the azimuthal velocity of the LSR (which equals the circular velocity if the disk is circular) is $v_y = 235 \pm 16$ km s$^{-1}$. This result is consistent with both
the traditional IAU value of 220 km s$^{-1}$ as well as some more recent methods that give somewhat larger results \citep[e.g.][]{2011MNRAS.414.2446M, 2012ApJ...759..131B}. While this new measurement does not help to resolve the controversy on the exact value of the solar motion, it does provide an independent consistency check on the SMHASH distances.

\section{Implications for the Milky Way's Mass}
\label{sec:mass}

Orbit fitting is known to introduce systematic biases in potential measures \citep{2011MNRAS.413.1852E,2013MNRAS.433.1813S,2013MNRAS.436.2386L}. To investigate what effect this might have for the specific case of the Orphan Stream, we have created N--body models of the stream and `observed' them in such a way as to recreate the SMHASH dataset. We then apply an identical orbit fitting technique and compare with the simulation inputs. This method allows us to contextualize the results of our RRL observations in terms of the direction and size of systematic biases as well as compare them with earlier results.

Previous measurement of the Milky Way's mass using the Orphan Stream found that the best--fitting halo was a factor of $\sim2$ less massive inside 60 kpc \citep[$2.74 \times 10^{11} \mathrm{M_{\odot}}$,][]{2010ApJ...711...32N} than contemporary models using other techniques, such as fitting Sagittarius Stream data \citep[$4.7 \times 10^{11} \mathrm{M_{\odot}}$,][]{2005ApJ...619..807L} or the velocity distribution of field BHB stars \citep[$4.0 \times 10^{11} \mathrm{M_{\odot}}$,][]{2008ApJ...684.1143X}. A complete summary of mass estimates is outside the scope of this work; the review of \cite{2016ARA&A..54..529B} provides an overview. However, the \cite{2010ApJ...711...32N} measurement remains below all published estimates and recent results reach masses only as low as about $3.2 \times 10^{11} \mathrm{M_{\odot}}$ \citep{2014MNRAS.445.3788G}.

\subsection{Creating and observing mock data sets}
\label{sec:nbody}

We use the self--consistent field method \citep[SCF, ][]{1992ApJ...386..375H}, which represents the gravitational potential of the disrupting satellite as a basis function expansion, to create a series of N--body simulations designed to reasonably mimic the observed Orphan Stream. The single--component, dark matter only Orphan progenitor is implemented as a Navarro--Frenk--White \citep[NFW,][]{1997ApJ...490..493N} distribution with $10^5$ particles. The particles are instantiated out to 35 scale radii and so the model's total mass differs from the virial mass; in the following we report the corresponding virial mass to avoid confusion. All simulations have the same mean density inside the scale radius, which results in tides unbinding them at approximately the same time. This allows the separation of effects due to the time of disruption and passive evolution. The density scaling is set such that the halo with a virial mass of $\mathrm{10^{9}\ M_\odot}$ has a scale radius of $0.75$ kpc although the results are not particularly sensitive to this choice. 

We chose the orbit and potential model to be precisely that of \citet{2010ApJ...711...32N}'s Model 5: that is, an orbit initialized from the phase space coordinate with Heliocentric position $(l,b,R)=(218^{\circ},53.5^{\circ},28.6\ \mathrm{kpc})$ and Galactocentric velocity $(v_x,v_y,v_z)=(−156, 79, 107) \ \mathrm{km\ s^{-1}}$ moving in a logarithmic potential model (Equations~\ref{eq:disk}-\ref{eq:halo}) with the one unspecified parameter $v_{\rm halo}$ set to $73\ \mathrm{km\ s^{-1}}$. The orbit is integrated backwards in time to find the phase space coordinate of the 3rd apocenter, 4.8 Gyr ago. When the satellite is near apocenter the hosts' tidal field is at its weakest, so beginning the simulation here minimizes artificial gravitational shocking. After relaxing in isolation the host potential is turned on over 10 internal dynamical times, the particle distribution is inserted, and the satellite is evolved to the present day. We assume that the current position of the progenitor is at the overdensity identified by \cite{2015ApJ...812L..26G}, $l \approx 268.7^\circ$, so the simulation ends at that point.

To produce synthetic observations that approximate those of the SMHASH RRL, we first select the particles below the tenth percentile in initial internal binding energy. These are tagged as stars. This simple strategy has been shown to reproduce the observed properties of Local Group dwarf galaxies in semianalytic models \citep{2005ApJ...635..931B} and create stellar haloes with realistic properties in simulations of Milky Way--like galaxies with cosmological infall \citep{2008MNRAS.391...14D,2010MNRAS.406..744C}. From this subset we choose at random 30 particles that match the selection criteria used in \cite{2013ApJ...776...26S}, namely Galactic longitude $260^\circ > l > 160^\circ$, Orphan latitude $4^\circ > B > -4^\circ$, and Galactic standard of rest velocity $v_{gsr} > 40\ \mathrm{km\ s^{-1}}$. Since the particle positions and velocities are precisely known, we introduce `observational' uncertainties by adding a random velocity drawn from a Gaussian of width $15\ \mathrm{km\ s^{-1}}$ to each particle's heliocentric velocity. Similarly, the selected particles are scattered in heliocentric distance according to the $2.5\%$ relative uncertainty demonstrated in Figure~\ref{fig:unc}. These same values are retained as uncertainties to be fed into the orbit fitting algorithm as well.

\begin{figure}
\includegraphics[width=\linewidth]{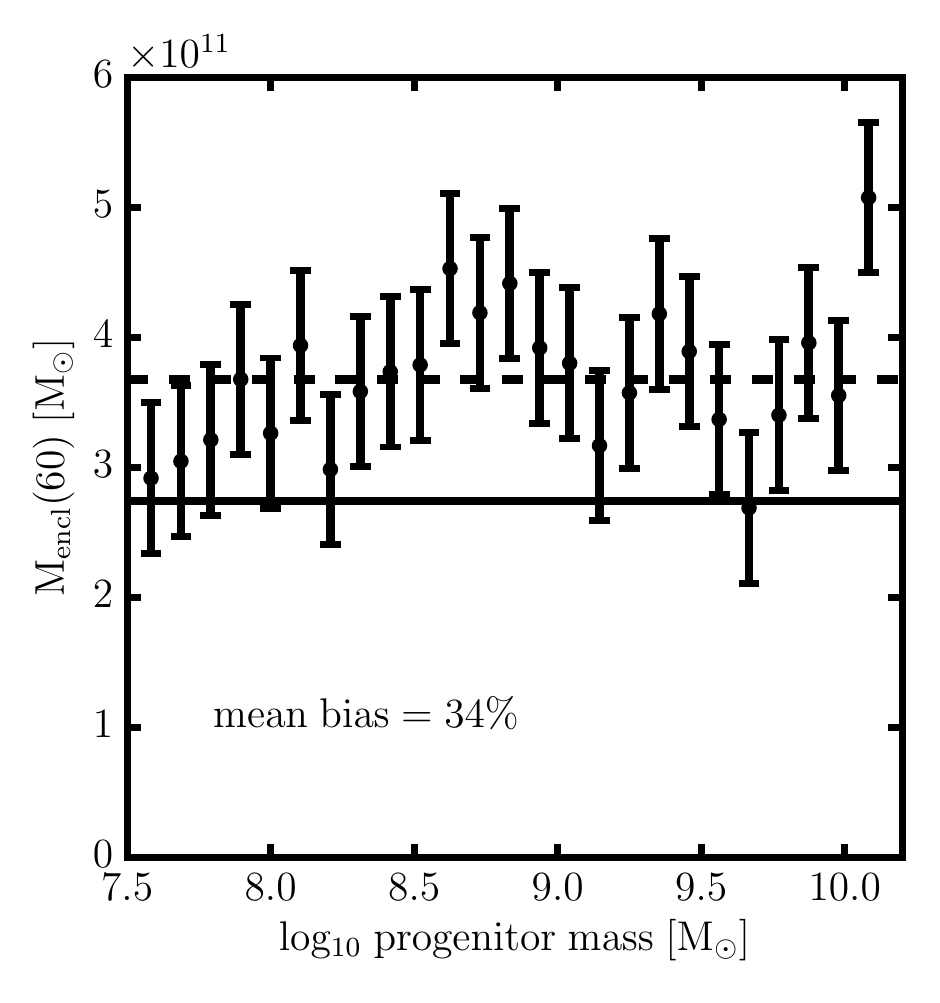}
\caption{\label{fig:simmass} Bias in the best--fitting host halo's enclosed mass, calculated from $v_{\rm halo}$, as a function of the initial halo mass of the progenitor satellite. The black horizontal line represents the true value in the model potential, while the points illustrate the posterior distribution of the fitted  $v_{\rm halo}$ in the N--body simulations. The dashed line is the mean fitted mass, which is greater than the true value by 34\%.} 
\end{figure}

\subsection{Biases in orbit fitting}
\label{sec:bias}
The problems associated with assuming stars in a tidal stream follow a single orbit are conceptually simplified when considering the Orphan Stream since we observe only the leading tail. In this case, stars farther from the satellite -- towards apocenter -- have lower total energy; their individual orbits turn around at smaller Galactocentric radii than the progenitor's does. Thus, orbits matched to the stream's path are tracing both the loss of kinetic energy to the gravitational potential as well as an additional loss determined by the total energy gradient of stars along the stream. Since the latter is not modelled in orbit fitting, the potential needs to be deeper at fixed radius to compensate for this `extra' loss, leading to an inflated mass estimate.

Figure~\ref{fig:simmass} illustrates the systematic error in inferred mass introduced by this effect. Despite the fact that each simulation was run in a potential with ${\rm M_{encl}(60\ kpc) = 2.7 \times 10^{11}\ M_\odot}$, the median value of the marginalized posterior distributions of $v_{\rm halo}$ generate an estimate $\sim 33$ per cent more massive. The bias is nearly independent of satellite mass, which matches theoretical expectations \citep{2013MNRAS.433.1813S}. To our knowledge this is the first time that the bias in mass enclosed due to orbit fitting has been quantified in a scenario that replicates an observed system. The magnitude of the effect likely depends on the details of the potential model but the direction should not -- the fitting algorithm will always prefer haloes that are more massive than are correct. For this reason we report the value measured for the Milky Way as only an upper limit.

We also note that the already low enclosed mass measurement of \cite{2010ApJ...711...32N} should also be affected by this systematic error since the approximation is the same despite their different fitting technique. If the magnitude of the bias is identical then the corrected mass enclosed is approximately ${\rm 1.8 \times 10^{11}\ M_\odot}$, slightly more than half that found by \cite{2014MNRAS.445.3788G}. Models with such small enclosed masses may have difficulty matching other observables such as the circular velocity of the Sun.

\section{The Orphan progenitor}
\label{sec:prog}

In the previous section we were concerned primarily with the model parameters that describe the phase space position of the orbits and the shape of the potential. Now we focus on the internal structure of the stream, characterized by the widths $\delta_{B},\ \delta_{v_{r}}$, and $\delta_{DM}$. For a particular progenitor orbit the spatial and velocity scales of the stream stars vary with the satellite--to--host mass ratio as $(m/M)^{1/3}$ \citep{1998ApJ...495..297J,1999MNRAS.307..495H,2001ApJ...557..137J}; therefore the $\delta_i$ contain information about the progenitor system. To first order this is the mass when the stars are unbound, however it may be possible to recover the satellite's central density distribution which also imprints itself on the stream \citep{2015MNRAS.449L..46E}.

\begin{figure}
\includegraphics[width=\linewidth]{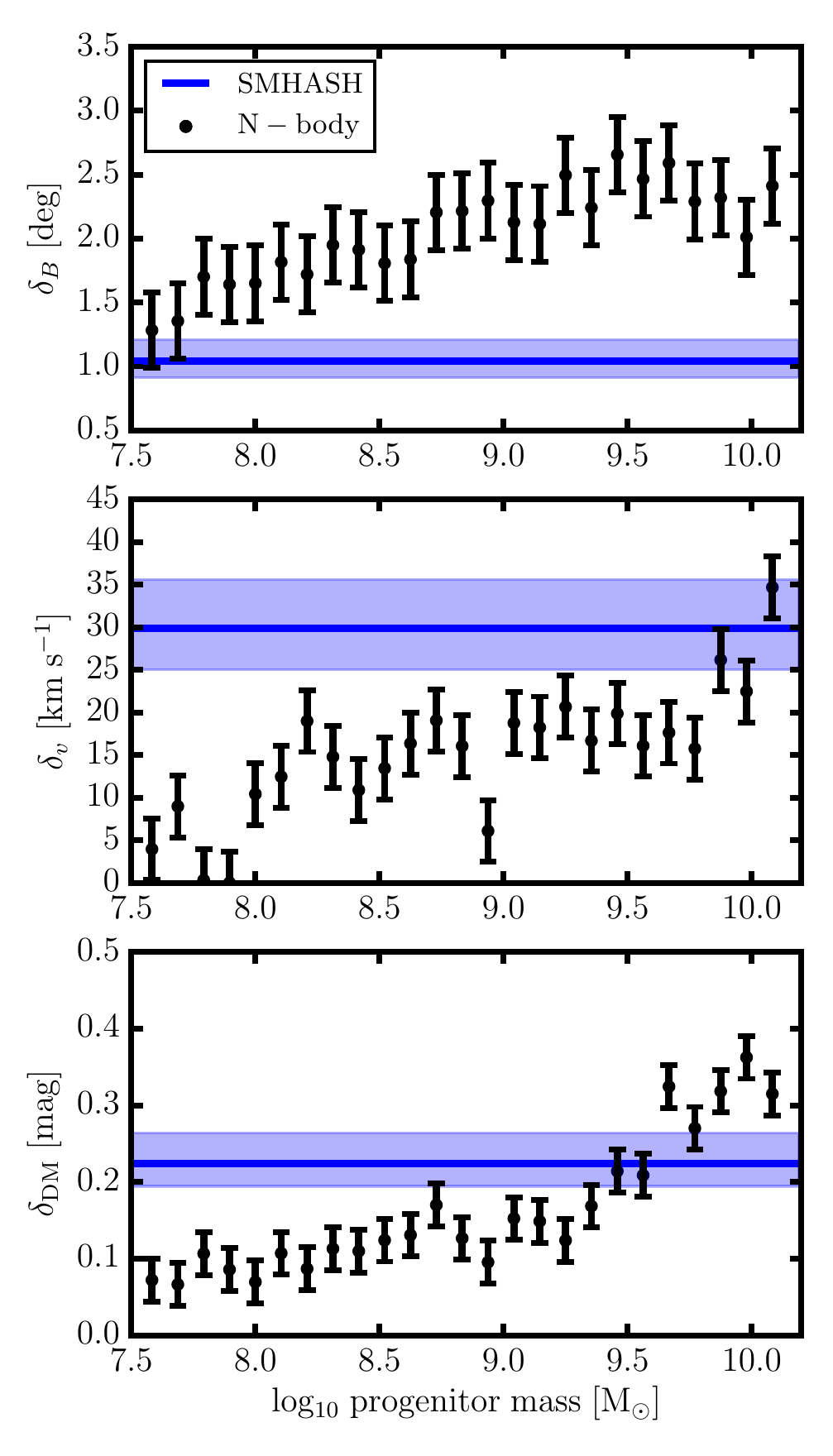}
\caption{\label{fig:simstats} Fitted values of width on the sky (top), velocity dispersion (center), and line--of--sight depth (bottom) for a set of N--body models of the Orphan Stream (black points) as a function of model satellite mass, compared to the same quantities as measured for the SMHASH Orphan data (blue region).} 
\end{figure}

Figure~\ref{fig:simstats} shows the effect of satellite mass on the simulated streams' structural parameters. In each panel the horizontal blue lines illustrate the values measured from the SMHASH data while the black points show the same quantities found after applying the same orbit fitting algorithm to N--body simulations of varying initial satellite halo masses. The mass range shown, from $3.8 \times 10^{7}$ to $1.2 \times 10^{10}\ {\rm M_\odot}$, captures dwarf galaxies from the ultrafaints to a few times less massive than the Small Magellanic Cloud \citep{2010MNRAS.404.1111G}.

First we consider the stream's width on the sky, $\delta_B$, plotted in the upper panel. The measured value $\delta_B = 1^\circ$ appears at a glance to be most consistent with the lowest--mass simulations, indicating that $\mathrm{M_{Orphan} \approx 10^8\ M_\odot}$. However, the selection of RR Lyrae stars for spectroscopic follow--up in in the SMHASH precursor catalogues is non--uniform and appears to be weighted significantly towards stars that are nearer the stream centre (e.g., of the stars with $2^\circ<B<4^\circ$, 3 have spectra and 11 do not). The observed $\delta_B$ is therefore unlikely to be representative of the true distribution. An alternative approach is to look at studies of Orphan's main sequence population; since our synthetic RRL are selected at random from the star particles, they represent any other stellar population just as well under the assumption that Orphan was originally well mixed. Belokurov et al. (2007) found that the stream has a full--width half--max of around $2^\circ$, which is comparable to the SMHASH RRL $\delta_B = 1^\circ$.  However, \cite{2008MNRAS.389.1391S} showed that the observed stream width may be truncated by confusion with the Galactic background and that streams as wide as $15^\circ$ could be hidden in the data. We therefore take $\delta_B$ as measured in SMHASH as a lower limit on acceptable values in the N--body simulations, indicating $\mathrm{M_{Orphan} \gtrsim 10^8\ M_\odot}$.

Next, we consider the velocity dispersion $\delta_{v_{r}}$, shown in the middle panel of Figure~\ref{fig:simstats}. It is clear that our model fits cannot reproduce the observed velocity dispersion except in the case of the highest mass progenitors. In fact, the true dispersion is buried by the simulated velocity errors for the lower mass models, resulting in a flat profile across much of the mass range. To obtain the 30 km s$^{-1}$ required to match the $\delta_{v_{r}}$ fit to the \citep{2013ApJ...776...26S} velocities would require a satellite of mass $\gtrsim 10^{10}\ M_\odot$. Such a progenitor seems unlikely given Orphan's luminosity and metallicity as well as the other structural parameters. 
In addition, \cite{2010ApJ...711...32N} measured the velocity dispersion of Orphan's BHB stars and found $\sigma_v = 8-13$ km s$^{-1}$ at various points along the stream; 
similarly, the K-giants surveyed by \cite{2013ApJ...764...39C} have a velocity dispersion of $6.5 \pm 7.0$ km s$^{-1}$. Values in the $5-15$ km s$^{-1}$ range are consistent with a wide variety of N--body models.
We note that obtaining systemic velocities for RRL requires subtraction of the stars' atmospheric velocities as they pulsate. The velocity variation of spectral lines over a single cycle can approach 100 km s$^{-1}$ \citep[e.g.][]{2011AJ....141....6P}, so if even a fraction remains it could explain this discrepancy. Due to this concerns we place lower weight on $\delta_{v_{r}}$ as a constraint and consider it as only an upper limit on progenitor mass.

Finally, the bottom panel of Figure~\ref{fig:simstats} shows the trend of line--of--sight depth in distance modulus,  $\delta_{DM}$, as a function of progenitor mass. Of our measurements this dimension provides the most confident constraint on the Orphan progenitor. A line fit to the apparently linear behaviour of the models above $\mathrm{10^9\ M_\odot}$ shows that an initial mass $\mathrm{M_{Orphan} \approx 3.2 \times 10^9\ M_\odot}$ best reproduces the observed depth of 0.224 mag. At high satellite mass the stream begins fanning out near apocenter due to azimuthal precession of the orbits, leading to larger depths and increased dependence of measured parameters on the selection of simulation particles as RRL.

Taken as a whole, the structure of the stream suggests a progenitor with initial halo mass of several times ${\rm 10^9\ M_\odot}$.  That value is in good agreement with the Local Group dwarf spheroidals, who seem to live in haloes in this range \citep{2008ApJ...672..904P, 2012MNRAS.422.1203B, 2016arXiv160706479F} and provides further evidence that Orphan is indeed a disrupted dwarf spheroidal galaxy. Satellite mass measurements obtained in this way are naturally potential--dependent since the stream structure is sensitive principally to the mass ratio. While the average $v_{\rm halo}$ fit in the N--body models is well matched to that of SMHASH we cannot say with confidence that the bias will be identical. Using any literature value for the Milky Way's mass will vary this result by less than a factor of 2, surely less than the systematic uncertainty in this simple method. 

\section{Summary}
\label{sec:concl}

This work presents {\it Spitzer Space Telescope} observations of 32 candidate Orphan Stream RR Lyrae stars as part of the {\it Spitzer} Merger History and Shape of the Galactic Halo (SMHASH) program. Using a theoretical period--luminosity--metallicity relation at $\mathrm{3.6 \mu m}$ in conjunction with archival data we have obtained distances to individual stars with 2.5\% relative uncertainties, a factor of two better than the previous state of the art. We find that the stream extends to approximately 50 kpc in heliocentric distance within the survey footprint and have resolved its large line--of--sight depth of approximately 8 kpc as it approaches apocenter.

Using a Markov Chain Monte Carlo orbit fitting algorithm, we find that the SMHASH data are consistent with a more massive Milky Way halo than indicated by previous work using same stream and a similar technique. By comparing with N--body simulations of dwarf galaxy tidal disruptions, we find that orbits fit to the available Orphan data are biased to high masses, suggesting that our measurement is an upper limit (and in good agreement with other modern methodologies). While proper motion measurements seem to provide significant leverage on the Milky Way's halo, our potential model is apparently too rigid to take advantage of the full phase space information. Integrating six-dimensional constraints are a promising avenue for future work.

By examining the structure of the stream -- namely its line--of--sight depth, velocity dispersion, and width on the sky -- we find that a satellite galaxy with an initial halo mass $\mathrm{M_{Orphan} \approx 3.2 \times 10^9\ M_\odot}$ best reproduces the SMHASH data. In combination with the integrated luminosity of the stream, this indicates that the progenitor was likely comparable to the Milky Way's eight classical dwarf spheroidals.

The SMHASH RR Lyrae star distances are fertile ground for further detailed study of the Orphan Stream. The investigations presented here represent only a first step towards understanding this surprisingly complex object. Future work, including implementing sophisticated potential measuring techniques and leveraging additional data from the {\it Gaia} mission and others, promises to improve our knowledge of the Milky Way and its satellite system.

\section*{Acknowledgements}
DH thanks Peter Stetson, Sheila Kannappan, and Vasily Belokurov for helpful discussions. This work is based on observations made with the {\it Spitzer Space Telescope}, which is operated by the Jet Propulsion Laboratory, California Institute of Technology under a contract with NASA. DH and KVJ acknowledge support on various aspects of this project from NASA through subcontract JPL 1558281 and ATP grant NNX15AK78G, as well as from the NSF through the grant AST-1614743. The Space Telescope Science Institute (STScI) co--authors acknowledge NASA support through a grant for HST program GO-13443 from STScI, which is operated by the Association of Universities for Research in Astronomy (AURA), Inc., under NASA contract NAS5-26555. DH acknowledges the use of the Shared Research Computing Facility at Columbia University. This work made use of Matplotlib \citep{Hunter:2007}, SciPy \citep{scipy}, Astropy \citep{2013A&A...558A..33A}, and the Astropy--affiliated Gala package \citep{gala}.

\bibliographystyle{mnras}
\bibliography{orphanbib}

\begin{thebibliography}{}
\makeatletter
\relax
\def\mn@urlcharsother{\let\do\@makeother \do\$\do\&\do\#\do\^\do\_\do\%\do\~}
\def\mn@doi{\begingroup\mn@urlcharsother \@ifnextchar [ {\mn@doi@}
  {\mn@doi@[]}}
\def\mn@doi@[#1]#2{\def\@tempa{#1}\ifx\@tempa\@empty \href
  {http://dx.doi.org/#2} {doi:#2}\else \href {http://dx.doi.org/#2} {#1}\fi
  \endgroup}
\def\mn@eprint#1#2{\mn@eprint@#1:#2::\@nil}
\def\mn@eprint@arXiv#1{\href {http://arxiv.org/abs/#1} {{\tt arXiv:#1}}}
\def\mn@eprint@dblp#1{\href {http://dblp.uni-trier.de/rec/bibtex/#1.xml}
  {dblp:#1}}
\def\mn@eprint@#1:#2:#3:#4\@nil{\def\@tempa {#1}\def\@tempb {#2}\def\@tempc
  {#3}\ifx \@tempc \@empty \let \@tempc \@tempb \let \@tempb \@tempa \fi \ifx
  \@tempb \@empty \def\@tempb {arXiv}\fi \@ifundefined
  {mn@eprint@\@tempb}{\@tempb:\@tempc}{\expandafter \expandafter \csname
  mn@eprint@\@tempb\endcsname \expandafter{\@tempc}}}

\bibitem[\protect\citeauthoryear{{Amorisco}}{{Amorisco}}{2015}]{2015MNRAS.450..575A}
{Amorisco} N.~C.,  2015, \mn@doi [\mnras] {10.1093/mnras/stv648}, \href
  {http://adsabs.harvard.edu/abs/2015MNRAS.450..575A} {450, 575}

\bibitem[\protect\citeauthoryear{{Astropy Collaboration} et~al.,}{{Astropy
  Collaboration} et~al.}{2013}]{2013A&A...558A..33A}
{Astropy Collaboration} et~al., 2013, \mn@doi [\aap]
  {10.1051/0004-6361/201322068}, \href
  {http://adsabs.harvard.edu/abs/2013A%26A...558A..33A} {558, A33}

\bibitem[\protect\citeauthoryear{{Belokurov} et~al.,}{{Belokurov}
  et~al.}{2006}]{2006ApJ...642L.137B}
{Belokurov} V.,  et~al., 2006, \mn@doi [\apjl] {10.1086/504797}, \href
  {http://adsabs.harvard.edu/abs/2006ApJ...642L.137B} {642, L137}

\bibitem[\protect\citeauthoryear{{Belokurov} et~al.,}{{Belokurov}
  et~al.}{2007}]{2007ApJ...658..337B}
{Belokurov} V.,  et~al., 2007, \mn@doi [\apj] {10.1086/511302}, \href
  {http://adsabs.harvard.edu/abs/2007ApJ...658..337B} {658, 337}

\bibitem[\protect\citeauthoryear{{Bersier} \& {Wood}}{{Bersier} \&
  {Wood}}{2002}]{2002AJ....123..840B}
{Bersier} D.,  {Wood} P.~R.,  2002, \mn@doi [\aj] {10.1086/338315}, \href
  {http://adsabs.harvard.edu/abs/2002AJ....123..840B} {123, 840}

\bibitem[\protect\citeauthoryear{{Bland-Hawthorn} \&
  {Gerhard}}{{Bland-Hawthorn} \& {Gerhard}}{2016}]{2016ARA&A..54..529B}
{Bland-Hawthorn} J.,  {Gerhard} O.,  2016, \mn@doi [\araa]
  {10.1146/annurev-astro-081915-023441}, \href
  {http://adsabs.harvard.edu/abs/2016ARA%26A..54..529B} {54, 529}

\bibitem[\protect\citeauthoryear{{Bono}, {Caputo}, {Castellani}, {Marconi}  \&
  {Storm}}{{Bono} et~al.}{2001}]{2001MNRAS.326.1183B}
{Bono} G.,  {Caputo} F.,  {Castellani} V.,  {Marconi} M.,   {Storm} J.,  2001,
  \mn@doi [\mnras] {10.1046/j.1365-8711.2001.04655.x}, \href
  {http://adsabs.harvard.edu/abs/2001MNRAS.326.1183B} {326, 1183}

\bibitem[\protect\citeauthoryear{{Bono}, {Caputo}, {Castellani}, {Marconi},
  {Storm}  \& {Degl'Innocenti}}{{Bono} et~al.}{2003}]{2003MNRAS.344.1097B}
{Bono} G.,  {Caputo} F.,  {Castellani} V.,  {Marconi} M.,  {Storm} J.,
  {Degl'Innocenti} S.,  2003, \mn@doi [\mnras]
  {10.1046/j.1365-8711.2003.06878.x}, \href
  {http://adsabs.harvard.edu/abs/2003MNRAS.344.1097B} {344, 1097}

\bibitem[\protect\citeauthoryear{{Bovy} et~al.,}{{Bovy}
  et~al.}{2012}]{2012ApJ...759..131B}
{Bovy} J.,  et~al., 2012, \mn@doi [\apj] {10.1088/0004-637X/759/2/131}, \href
  {http://adsabs.harvard.edu/abs/2012ApJ...759..131B} {759, 131}

\bibitem[\protect\citeauthoryear{{Bovy}, {Bahmanyar}, {Fritz}  \&
  {Kallivayalil}}{{Bovy} et~al.}{2016}]{2016ApJ...833...31B}
{Bovy} J.,  {Bahmanyar} A.,  {Fritz} T.~K.,   {Kallivayalil} N.,  2016, \mn@doi
  [\apj] {10.3847/1538-4357/833/1/31}, \href
  {http://adsabs.harvard.edu/abs/2016ApJ...833...31B} {833, 31}

\bibitem[\protect\citeauthoryear{{Boylan-Kolchin}, {Bullock}  \&
  {Kaplinghat}}{{Boylan-Kolchin} et~al.}{2012}]{2012MNRAS.422.1203B}
{Boylan-Kolchin} M.,  {Bullock} J.~S.,   {Kaplinghat} M.,  2012, \mn@doi
  [\mnras] {10.1111/j.1365-2966.2012.20695.x}, \href
  {http://adsabs.harvard.edu/abs/2012MNRAS.422.1203B} {422, 1203}

\bibitem[\protect\citeauthoryear{{Braga} et~al.,}{{Braga}
  et~al.}{2015}]{2015ApJ...799..165B}
{Braga} V.~F.,  et~al., 2015, \mn@doi [\apj] {10.1088/0004-637X/799/2/165},
  \href {http://adsabs.harvard.edu/abs/2015ApJ...799..165B} {799, 165}

\bibitem[\protect\citeauthoryear{{Bullock} \& {Johnston}}{{Bullock} \&
  {Johnston}}{2005}]{2005ApJ...635..931B}
{Bullock} J.~S.,  {Johnston} K.~V.,  2005, \mn@doi [\apj] {10.1086/497422},
  \href {http://adsabs.harvard.edu/abs/2005ApJ...635..931B} {635, 931}

\bibitem[\protect\citeauthoryear{{Bullock}, {Kravtsov}  \&
  {Weinberg}}{{Bullock} et~al.}{2001}]{2001ApJ...548...33B}
{Bullock} J.~S.,  {Kravtsov} A.~V.,   {Weinberg} D.~H.,  2001, \mn@doi [\apj]
  {10.1086/318681}, \href {http://adsabs.harvard.edu/abs/2001ApJ...548...33B}
  {548, 33}

\bibitem[\protect\citeauthoryear{{Cacciari} \& {Clementini}}{{Cacciari} \&
  {Clementini}}{2003}]{2003LNP...635..105C}
{Cacciari} C.,  {Clementini} G.,  2003, in {Alloin} D.,  {Gieren} W.,  eds,
  Lecture Notes in Physics, Berlin Springer Verlag Vol. 635, Stellar Candles
  for the Extragalactic Distance Scale. pp 105--122 (\mn@eprint {}
  {astro-ph/0301550}), \mn@doi{10.1007/978-3-540-39882-0_6}

\bibitem[\protect\citeauthoryear{{Cardelli}, {Clayton}  \& {Mathis}}{{Cardelli}
  et~al.}{1989}]{1989ApJ...345..245C}
{Cardelli} J.~A.,  {Clayton} G.~C.,   {Mathis} J.~S.,  1989, \mn@doi [\apj]
  {10.1086/167900}, \href {http://adsabs.harvard.edu/abs/1989ApJ...345..245C}
  {345, 245}

\bibitem[\protect\citeauthoryear{{Carlin}, {Majewski}, {Casetti-Dinescu},
  {Law}, {Girard}  \& {Patterson}}{{Carlin} et~al.}{2012}]{2012ApJ...744...25C}
{Carlin} J.~L.,  {Majewski} S.~R.,  {Casetti-Dinescu} D.~I.,  {Law} D.~R.,
  {Girard} T.~M.,   {Patterson} R.~J.,  2012, \mn@doi [\apj]
  {10.1088/0004-637X/744/1/25}, \href
  {http://adsabs.harvard.edu/abs/2012ApJ...744...25C} {744, 25}

\bibitem[\protect\citeauthoryear{{Casey}, {Da Costa}, {Keller}  \&
  {Maunder}}{{Casey} et~al.}{2013}]{2013ApJ...764...39C}
{Casey} A.~R.,  {Da Costa} G.,  {Keller} S.~C.,   {Maunder} E.,  2013, \mn@doi
  [\apj] {10.1088/0004-637X/764/1/39}, \href
  {http://adsabs.harvard.edu/abs/2013ApJ...764...39C} {764, 39}

\bibitem[\protect\citeauthoryear{{Catelan}, {Pritzl}  \& {Smith}}{{Catelan}
  et~al.}{2004}]{2004ApJS..154..633C}
{Catelan} M.,  {Pritzl} B.~J.,   {Smith} H.~A.,  2004, \mn@doi [\apjs]
  {10.1086/422916}, \href {http://adsabs.harvard.edu/abs/2004ApJS..154..633C}
  {154, 633}

\bibitem[\protect\citeauthoryear{{Cohen}, {Sesar}, {Bahnolzer}, {He},
  {Kulkarni}, {Prince}, {Bellm}  \& {Laher}}{{Cohen}
  et~al.}{2017}]{2017arXiv171001276C}
{Cohen} J.,  {Sesar} B.,  {Bahnolzer} S.,  {He} K.,  {Kulkarni} S.~R.,
  {Prince} T.~A.,  {Bellm} E.,   {Laher} R.~R.,  2017, preprint, \href
  {http://adsabs.harvard.edu/abs/2017arXiv171001276C} {} (\mn@eprint {arXiv}
  {1710.01276})

\bibitem[\protect\citeauthoryear{{Cooper} et~al.,}{{Cooper}
  et~al.}{2010}]{2010MNRAS.406..744C}
{Cooper} A.~P.,  et~al., 2010, \mn@doi [\mnras]
  {10.1111/j.1365-2966.2010.16740.x}, \href
  {http://adsabs.harvard.edu/abs/2010MNRAS.406..744C} {406, 744}

\bibitem[\protect\citeauthoryear{{De Lucia} \& {Helmi}}{{De Lucia} \&
  {Helmi}}{2008}]{2008MNRAS.391...14D}
{De Lucia} G.,  {Helmi} A.,  2008, \mn@doi [\mnras]
  {10.1111/j.1365-2966.2008.13862.x}, \href
  {http://adsabs.harvard.edu/abs/2008MNRAS.391...14D} {391, 14}

\bibitem[\protect\citeauthoryear{{Drake} et~al.,}{{Drake}
  et~al.}{2009}]{2009ApJ...696..870D}
{Drake} A.~J.,  et~al., 2009, \mn@doi [\apj] {10.1088/0004-637X/696/1/870},
  \href {http://adsabs.harvard.edu/abs/2009ApJ...696..870D} {696, 870}

\bibitem[\protect\citeauthoryear{{Drake} et~al.,}{{Drake}
  et~al.}{2013}]{2013ApJ...763...32D}
{Drake} A.~J.,  et~al., 2013, \mn@doi [\apj] {10.1088/0004-637X/763/1/32},
  \href {http://adsabs.harvard.edu/abs/2013ApJ...763...32D} {763, 32}

\bibitem[\protect\citeauthoryear{{Errani}, {Pe{\~n}arrubia}  \&
  {Tormen}}{{Errani} et~al.}{2015}]{2015MNRAS.449L..46E}
{Errani} R.,  {Pe{\~n}arrubia} J.,   {Tormen} G.,  2015, \mn@doi [\mnras]
  {10.1093/mnrasl/slv012}, \href
  {http://adsabs.harvard.edu/abs/2015MNRAS.449L..46E} {449, L46}

\bibitem[\protect\citeauthoryear{{Eyre} \& {Binney}}{{Eyre} \&
  {Binney}}{2011}]{2011MNRAS.413.1852E}
{Eyre} A.,  {Binney} J.,  2011, \mn@doi [\mnras]
  {10.1111/j.1365-2966.2011.18270.x}, \href
  {http://adsabs.harvard.edu/abs/2011MNRAS.413.1852E} {413, 1852}

\bibitem[\protect\citeauthoryear{{Fakhouri}, {Ma}  \&
  {Boylan-Kolchin}}{{Fakhouri} et~al.}{2010}]{2010MNRAS.406.2267F}
{Fakhouri} O.,  {Ma} C.-P.,   {Boylan-Kolchin} M.,  2010, \mn@doi [\mnras]
  {10.1111/j.1365-2966.2010.16859.x}, \href
  {http://adsabs.harvard.edu/abs/2010MNRAS.406.2267F} {406, 2267}

\bibitem[\protect\citeauthoryear{{Fattahi}, {Navarro}, {Sawala}, {Frenk},
  {Sales}, {Oman}, {Schaller}  \& {Wang}}{{Fattahi}
  et~al.}{2016}]{2016arXiv160706479F}
{Fattahi} A.,  {Navarro} J.~F.,  {Sawala} T.,  {Frenk} C.~S.,  {Sales} L.~V.,
  {Oman} K.,  {Schaller} M.,   {Wang} J.,  2016, preprint, \href
  {http://adsabs.harvard.edu/abs/2016arXiv160706479F} {} (\mn@eprint {arXiv}
  {1607.06479})

\bibitem[\protect\citeauthoryear{{Fazio} et~al.,}{{Fazio}
  et~al.}{2004}]{2004ApJS..154...10F}
{Fazio} G.~G.,  et~al., 2004, \mn@doi [\apjs] {10.1086/422843}, \href
  {http://adsabs.harvard.edu/abs/2004ApJS..154...10F} {154, 10}

\bibitem[\protect\citeauthoryear{{Fiorentino} et~al.,}{{Fiorentino}
  et~al.}{2015}]{2015ApJ...798L..12F}
{Fiorentino} G.,  et~al., 2015, \mn@doi [\apjl] {10.1088/2041-8205/798/1/L12},
  \href {http://adsabs.harvard.edu/abs/2015ApJ...798L..12F} {798, L12}

\bibitem[\protect\citeauthoryear{Foreman-Mackey}{Foreman-Mackey}{2016}]{corner}
Foreman-Mackey D.,  2016, \mn@doi [The Journal of Open Source Software]
  {10.21105/joss.00024}, 24

\bibitem[\protect\citeauthoryear{{Foreman-Mackey}, {Hogg}, {Lang}  \&
  {Goodman}}{{Foreman-Mackey} et~al.}{2013}]{2013PASP..125..306F}
{Foreman-Mackey} D.,  {Hogg} D.~W.,  {Lang} D.,   {Goodman} J.,  2013, \mn@doi
  [\pasp] {10.1086/670067}, \href
  {http://adsabs.harvard.edu/abs/2013PASP..125..306F} {125, 306}

\bibitem[\protect\citeauthoryear{{Freedman}, {Madore}, {Scowcroft}, {Burns},
  {Monson}, {Persson}, {Seibert}  \& {Rigby}}{{Freedman}
  et~al.}{2012}]{2012ApJ...758...24F}
{Freedman} W.~L.,  {Madore} B.~F.,  {Scowcroft} V.,  {Burns} C.,  {Monson} A.,
  {Persson} S.~E.,  {Seibert} M.,   {Rigby} J.,  2012, \mn@doi [\apj]
  {10.1088/0004-637X/758/1/24}, \href
  {http://adsabs.harvard.edu/abs/2012ApJ...758...24F} {758, 24}

\bibitem[\protect\citeauthoryear{{Freeman} \& {Bland-Hawthorn}}{{Freeman} \&
  {Bland-Hawthorn}}{2002}]{2002ARA&A..40..487F}
{Freeman} K.,  {Bland-Hawthorn} J.,  2002, \mn@doi [\araa]
  {10.1146/annurev.astro.40.060401.093840}, \href
  {http://adsabs.harvard.edu/abs/2002ARA%26A..40..487F} {40, 487}

\bibitem[\protect\citeauthoryear{{Fritz} \& {Kallivayalil}}{{Fritz} \&
  {Kallivayalil}}{2015}]{2015ApJ...811..123F}
{Fritz} T.~K.,  {Kallivayalil} N.,  2015, \mn@doi [\apj]
  {10.1088/0004-637X/811/2/123}, \href
  {http://adsabs.harvard.edu/abs/2015ApJ...811..123F} {811, 123}

\bibitem[\protect\citeauthoryear{{Gibbons}, {Belokurov}  \& {Evans}}{{Gibbons}
  et~al.}{2014}]{2014MNRAS.445.3788G}
{Gibbons} S.~L.~J.,  {Belokurov} V.,   {Evans} N.~W.,  2014, \mn@doi [\mnras]
  {10.1093/mnras/stu1986}, \href
  {http://adsabs.harvard.edu/abs/2014MNRAS.445.3788G} {445, 3788}

\bibitem[\protect\citeauthoryear{{Gibbons}, {Belokurov}  \& {Evans}}{{Gibbons}
  et~al.}{2017}]{2017MNRAS.464..794G}
{Gibbons} S.~L.~J.,  {Belokurov} V.,   {Evans} N.~W.,  2017, \mn@doi [\mnras]
  {10.1093/mnras/stw2328}, \href
  {http://adsabs.harvard.edu/abs/2017MNRAS.464..794G} {464, 794}

\bibitem[\protect\citeauthoryear{{Gillessen}, {Eisenhauer}, {Fritz}, {Bartko},
  {Dodds-Eden}, {Pfuhl}, {Ott}  \& {Genzel}}{{Gillessen}
  et~al.}{2009}]{2009ApJ...707L.114G}
{Gillessen} S.,  {Eisenhauer} F.,  {Fritz} T.~K.,  {Bartko} H.,  {Dodds-Eden}
  K.,  {Pfuhl} O.,  {Ott} T.,   {Genzel} R.,  2009, \mn@doi [\apjl]
  {10.1088/0004-637X/707/2/L114}, \href
  {http://adsabs.harvard.edu/abs/2009ApJ...707L.114G} {707, L114}

\bibitem[\protect\citeauthoryear{{Goodman} \& {Weare}}{{Goodman} \&
  {Weare}}{2010}]{GW2010}
{Goodman} J.,  {Weare} J.,  2010, Comm. App. Math. Comp. Sci., 5, 65

\bibitem[\protect\citeauthoryear{{Grillmair}}{{Grillmair}}{2006}]{2006ApJ...645L..37G}
{Grillmair} C.~J.,  2006, \mn@doi [\apjl] {10.1086/505863}, \href
  {http://adsabs.harvard.edu/abs/2006ApJ...645L..37G} {645, L37}

\bibitem[\protect\citeauthoryear{{Grillmair}, {Hetherington}, {Carlberg}  \&
  {Willman}}{{Grillmair} et~al.}{2015}]{2015ApJ...812L..26G}
{Grillmair} C.~J.,  {Hetherington} L.,  {Carlberg} R.~G.,   {Willman} B.,
  2015, \mn@doi [\apjl] {10.1088/2041-8205/812/2/L26}, \href
  {http://adsabs.harvard.edu/abs/2015ApJ...812L..26G} {812, L26}

\bibitem[\protect\citeauthoryear{{Guo}, {White}, {Li}  \&
  {Boylan-Kolchin}}{{Guo} et~al.}{2010}]{2010MNRAS.404.1111G}
{Guo} Q.,  {White} S.,  {Li} C.,   {Boylan-Kolchin} M.,  2010, \mn@doi [\mnras]
  {10.1111/j.1365-2966.2010.16341.x}, \href
  {http://adsabs.harvard.edu/abs/2010MNRAS.404.1111G} {404, 1111}

\bibitem[\protect\citeauthoryear{{Helmi} \& {White}}{{Helmi} \&
  {White}}{1999}]{1999MNRAS.307..495H}
{Helmi} A.,  {White} S.~D.~M.,  1999, \mn@doi [\mnras]
  {10.1046/j.1365-8711.1999.02616.x}, \href
  {http://adsabs.harvard.edu/abs/1999MNRAS.307..495H} {307, 495}

\bibitem[\protect\citeauthoryear{{Hendel} \& {Johnston}}{{Hendel} \&
  {Johnston}}{2015}]{2015MNRAS.454.2472H}
{Hendel} D.,  {Johnston} K.~V.,  2015, \mn@doi [\mnras]
  {10.1093/mnras/stv2035}, \href
  {http://adsabs.harvard.edu/abs/2015MNRAS.454.2472H} {454, 2472}

\bibitem[\protect\citeauthoryear{{Hernitschek} et~al.,}{{Hernitschek}
  et~al.}{2017}]{2017arXiv171009436H}
{Hernitschek} N.,  et~al., 2017, preprint, \href
  {http://adsabs.harvard.edu/abs/2017arXiv171009436H} {} (\mn@eprint {arXiv}
  {1710.09436})

\bibitem[\protect\citeauthoryear{{Hernquist}}{{Hernquist}}{1990}]{1990ApJ...356..359H}
{Hernquist} L.,  1990, \mn@doi [\apj] {10.1086/168845}, \href
  {http://adsabs.harvard.edu/abs/1990ApJ...356..359H} {356, 359}

\bibitem[\protect\citeauthoryear{{Hernquist} \& {Ostriker}}{{Hernquist} \&
  {Ostriker}}{1992}]{1992ApJ...386..375H}
{Hernquist} L.,  {Ostriker} J.~P.,  1992, \mn@doi [\apj] {10.1086/171025},
  \href {http://adsabs.harvard.edu/abs/1992ApJ...386..375H} {386, 375}

\bibitem[\protect\citeauthoryear{Hunter}{Hunter}{2007}]{Hunter:2007}
Hunter J.~D.,  2007, \mn@doi [Computing In Science \& Engineering]
  {10.1109/MCSE.2007.55}, 9, 90

\bibitem[\protect\citeauthoryear{{Indebetouw} et~al.,}{{Indebetouw}
  et~al.}{2005}]{2005ApJ...619..931I}
{Indebetouw} R.,  et~al., 2005, \mn@doi [\apj] {10.1086/426679}, \href
  {http://adsabs.harvard.edu/abs/2005ApJ...619..931I} {619, 931}

\bibitem[\protect\citeauthoryear{{Johnston}}{{Johnston}}{1998}]{1998ApJ...495..297J}
{Johnston} K.~V.,  1998, \mn@doi [\apj] {10.1086/305273}, \href
  {http://adsabs.harvard.edu/abs/1998ApJ...495..297J} {495, 297}

\bibitem[\protect\citeauthoryear{{Johnston}, {Hernquist}  \&
  {Bolte}}{{Johnston} et~al.}{1996}]{1996ApJ...465..278J}
{Johnston} K.~V.,  {Hernquist} L.,   {Bolte} M.,  1996, \mn@doi [\apj]
  {10.1086/177418}, \href {http://adsabs.harvard.edu/abs/1996ApJ...465..278J}
  {465, 278}

\bibitem[\protect\citeauthoryear{{Johnston}, {Sackett}  \&
  {Bullock}}{{Johnston} et~al.}{2001}]{2001ApJ...557..137J}
{Johnston} K.~V.,  {Sackett} P.~D.,   {Bullock} J.~S.,  2001, \mn@doi [\apj]
  {10.1086/321644}, \href {http://adsabs.harvard.edu/abs/2001ApJ...557..137J}
  {557, 137}

\bibitem[\protect\citeauthoryear{{Johnston}, {Bullock}, {Sharma}, {Font},
  {Robertson}  \& {Leitner}}{{Johnston} et~al.}{2008}]{2008ApJ...689..936J}
{Johnston} K.~V.,  {Bullock} J.~S.,  {Sharma} S.,  {Font} A.,  {Robertson}
  B.~E.,   {Leitner} S.~N.,  2008, \mn@doi [\apj] {10.1086/592228}, \href
  {http://adsabs.harvard.edu/abs/2008ApJ...689..936J} {689, 936}

\bibitem[\protect\citeauthoryear{{Johnston} et~al.,}{{Johnston}
  et~al.}{2013}]{2013sptz.prop10015J}
{Johnston} K.,  et~al., 2013, {SMASH: Spitzer Merger History and Shape of the
  Galactic Halo}, Spitzer Proposal

\bibitem[\protect\citeauthoryear{Jones, Oliphant, Peterson  et~al.}{Jones
  et~al.}{2001}]{scipy}
Jones E.,  Oliphant T.,  Peterson P.,   et~al., 2001, {SciPy}: Open source
  scientific tools for {Python}, \url {http://www.scipy.org/}

\bibitem[\protect\citeauthoryear{{Keller}, {Murphy}, {Prior}, {DaCosta}  \&
  {Schmidt}}{{Keller} et~al.}{2008}]{2008ApJ...678..851K}
{Keller} S.~C.,  {Murphy} S.,  {Prior} S.,  {DaCosta} G.,   {Schmidt} B.,
  2008, \mn@doi [\apj] {10.1086/526516}, \href
  {http://adsabs.harvard.edu/abs/2008ApJ...678..851K} {678, 851}

\bibitem[\protect\citeauthoryear{{Kirby}, {Cohen}, {Guhathakurta}, {Cheng},
  {Bullock}  \& {Gallazzi}}{{Kirby} et~al.}{2013}]{2013ApJ...779..102K}
{Kirby} E.~N.,  {Cohen} J.~G.,  {Guhathakurta} P.,  {Cheng} L.,  {Bullock}
  J.~S.,   {Gallazzi} A.,  2013, \mn@doi [\apj] {10.1088/0004-637X/779/2/102},
  \href {http://adsabs.harvard.edu/abs/2013ApJ...779..102K} {779, 102}

\bibitem[\protect\citeauthoryear{{Koposov}, {Rix}  \& {Hogg}}{{Koposov}
  et~al.}{2010}]{2010ApJ...712..260K}
{Koposov} S.~E.,  {Rix} H.-W.,   {Hogg} D.~W.,  2010, \mn@doi [\apj]
  {10.1088/0004-637X/712/1/260}, \href
  {http://adsabs.harvard.edu/abs/2010ApJ...712..260K} {712, 260}

\bibitem[\protect\citeauthoryear{{Koposov} et~al.,}{{Koposov}
  et~al.}{2012}]{2012ApJ...750...80K}
{Koposov} S.~E.,  et~al., 2012, \mn@doi [\apj] {10.1088/0004-637X/750/1/80},
  \href {http://adsabs.harvard.edu/abs/2012ApJ...750...80K} {750, 80}

\bibitem[\protect\citeauthoryear{{K{\"u}pper}, {Balbinot}, {Bonaca},
  {Johnston}, {Hogg}, {Kroupa}  \& {Santiago}}{{K{\"u}pper}
  et~al.}{2015}]{2015ApJ...803...80K}
{K{\"u}pper} A.~H.~W.,  {Balbinot} E.,  {Bonaca} A.,  {Johnston} K.~V.,  {Hogg}
  D.~W.,  {Kroupa} P.,   {Santiago} B.~X.,  2015, \mn@doi [\apj]
  {10.1088/0004-637X/803/2/80}, \href
  {http://adsabs.harvard.edu/abs/2015ApJ...803...80K} {803, 80}

\bibitem[\protect\citeauthoryear{{Law} \& {Majewski}}{{Law} \&
  {Majewski}}{2010}]{2010ApJ...714..229L}
{Law} D.~R.,  {Majewski} S.~R.,  2010, \mn@doi [\apj]
  {10.1088/0004-637X/714/1/229}, \href
  {http://adsabs.harvard.edu/abs/2010ApJ...714..229L} {714, 229}

\bibitem[\protect\citeauthoryear{{Law}, {Johnston}  \& {Majewski}}{{Law}
  et~al.}{2005}]{2005ApJ...619..807L}
{Law} D.~R.,  {Johnston} K.~V.,   {Majewski} S.~R.,  2005, \mn@doi [\apj]
  {10.1086/426779}, \href {http://adsabs.harvard.edu/abs/2005ApJ...619..807L}
  {619, 807}

\bibitem[\protect\citeauthoryear{{Law} et~al.,}{{Law}
  et~al.}{2009}]{2009PASP..121.1395L}
{Law} N.~M.,  et~al., 2009, \mn@doi [\pasp] {10.1086/648598}, \href
  {http://adsabs.harvard.edu/abs/2009PASP..121.1395L} {121, 1395}

\bibitem[\protect\citeauthoryear{{Layden}}{{Layden}}{1994}]{1994AJ....108.1016L}
{Layden} A.~C.,  1994, \mn@doi [\aj] {10.1086/117132}, \href
  {http://adsabs.harvard.edu/abs/1994AJ....108.1016L} {108, 1016}

\bibitem[\protect\citeauthoryear{{Longmore}, {Fernley}  \&
  {Jameson}}{{Longmore} et~al.}{1986}]{1986MNRAS.220..279L}
{Longmore} A.~J.,  {Fernley} J.~A.,   {Jameson} R.~F.,  1986, \mn@doi [\mnras]
  {10.1093/mnras/220.2.279}, \href
  {http://adsabs.harvard.edu/abs/1986MNRAS.220..279L} {220, 279}

\bibitem[\protect\citeauthoryear{{Lux}, {Read}, {Lake}  \& {Johnston}}{{Lux}
  et~al.}{2013}]{2013MNRAS.436.2386L}
{Lux} H.,  {Read} J.~I.,  {Lake} G.,   {Johnston} K.~V.,  2013, \mn@doi
  [\mnras] {10.1093/mnras/stt1744}, \href
  {http://adsabs.harvard.edu/abs/2013MNRAS.436.2386L} {436, 2386}

\bibitem[\protect\citeauthoryear{{Madore} et~al.,}{{Madore}
  et~al.}{2013}]{2013ApJ...776..135M}
{Madore} B.~F.,  et~al., 2013, \mn@doi [\apj] {10.1088/0004-637X/776/2/135},
  \href {http://adsabs.harvard.edu/abs/2013ApJ...776..135M} {776, 135}

\bibitem[\protect\citeauthoryear{{Majewski}, {Skrutskie}, {Weinberg}  \&
  {Ostheimer}}{{Majewski} et~al.}{2003}]{2003ApJ...599.1082M}
{Majewski} S.~R.,  {Skrutskie} M.~F.,  {Weinberg} M.~D.,   {Ostheimer} J.~C.,
  2003, \mn@doi [\apj] {10.1086/379504}, \href
  {http://adsabs.harvard.edu/abs/2003ApJ...599.1082M} {599, 1082}

\bibitem[\protect\citeauthoryear{{Makovoz} \& {Khan}}{{Makovoz} \&
  {Khan}}{2005}]{2005ASPC..347...81M}
{Makovoz} D.,  {Khan} I.,  2005, in {Shopbell} P.,  {Britton} M.,   {Ebert} R.,
   eds,  Astronomical Society of the Pacific Conference Series Vol. 347,
  Astronomical Data Analysis Software and Systems XIV. p.~81

\bibitem[\protect\citeauthoryear{{McMillan}}{{McMillan}}{2011}]{2011MNRAS.414.2446M}
{McMillan} P.~J.,  2011, \mn@doi [\mnras] {10.1111/j.1365-2966.2011.18564.x},
  \href {http://adsabs.harvard.edu/abs/2011MNRAS.414.2446M} {414, 2446}

\bibitem[\protect\citeauthoryear{{Miyamoto} \& {Nagai}}{{Miyamoto} \&
  {Nagai}}{1975}]{1975PASJ...27..533M}
{Miyamoto} M.,  {Nagai} R.,  1975, \pasj, \href
  {http://adsabs.harvard.edu/abs/1975PASJ...27..533M} {27, 533}

\bibitem[\protect\citeauthoryear{{Navarro}, {Frenk}  \& {White}}{{Navarro}
  et~al.}{1997}]{1997ApJ...490..493N}
{Navarro} J.~F.,  {Frenk} C.~S.,   {White} S.~D.~M.,  1997, \mn@doi [\apj]
  {10.1086/304888}, \href {http://adsabs.harvard.edu/abs/1997ApJ...490..493N}
  {490, 493}

\bibitem[\protect\citeauthoryear{{Neeley} et~al.,}{{Neeley}
  et~al.}{2015}]{2015ApJ...808...11N}
{Neeley} J.~R.,  et~al., 2015, \mn@doi [\apj] {10.1088/0004-637X/808/1/11},
  \href {http://adsabs.harvard.edu/abs/2015ApJ...808...11N} {808, 11}

\bibitem[\protect\citeauthoryear{{Neeley} et~al.,}{{Neeley}
  et~al.}{2017}]{2017ApJ...841...84N}
{Neeley} J.~R.,  et~al., 2017, \mn@doi [\apj] {10.3847/1538-4357/aa713d}, \href
  {http://adsabs.harvard.edu/abs/2017ApJ...841...84N} {841, 84}

\bibitem[\protect\citeauthoryear{{Newberg}, {Willett}, {Yanny}  \&
  {Xu}}{{Newberg} et~al.}{2010}]{2010ApJ...711...32N}
{Newberg} H.~J.,  {Willett} B.~A.,  {Yanny} B.,   {Xu} Y.,  2010, \mn@doi
  [\apj] {10.1088/0004-637X/711/1/32}, \href
  {http://adsabs.harvard.edu/abs/2010ApJ...711...32N} {711, 32}

\bibitem[\protect\citeauthoryear{{Pawlowski}, {Pflamm-Altenburg}  \&
  {Kroupa}}{{Pawlowski} et~al.}{2012}]{2012MNRAS.423.1109P}
{Pawlowski} M.~S.,  {Pflamm-Altenburg} J.,   {Kroupa} P.,  2012, \mn@doi
  [\mnras] {10.1111/j.1365-2966.2012.20937.x}, \href
  {http://adsabs.harvard.edu/abs/2012MNRAS.423.1109P} {423, 1109}

\bibitem[\protect\citeauthoryear{{Pe{\~n}arrubia}, {McConnachie}  \&
  {Navarro}}{{Pe{\~n}arrubia} et~al.}{2008}]{2008ApJ...672..904P}
{Pe{\~n}arrubia} J.,  {McConnachie} A.~W.,   {Navarro} J.~F.,  2008, \mn@doi
  [\apj] {10.1086/521543}, \href
  {http://adsabs.harvard.edu/abs/2008ApJ...672..904P} {672, 904}

\bibitem[\protect\citeauthoryear{{Pearson}, {K{\"u}pper}, {Johnston}  \&
  {Price-Whelan}}{{Pearson} et~al.}{2015}]{2015ApJ...799...28P}
{Pearson} S.,  {K{\"u}pper} A.~H.~W.,  {Johnston} K.~V.,   {Price-Whelan}
  A.~M.,  2015, \mn@doi [\apj] {10.1088/0004-637X/799/1/28}, \href
  {http://adsabs.harvard.edu/abs/2015ApJ...799...28P} {799, 28}

\bibitem[\protect\citeauthoryear{{Persson}, {Madore}, {Krzemi{\'n}ski},
  {Freedman}, {Roth}  \& {Murphy}}{{Persson}
  et~al.}{2004}]{2004AJ....128.2239P}
{Persson} S.~E.,  {Madore} B.~F.,  {Krzemi{\'n}ski} W.,  {Freedman} W.~L.,
  {Roth} M.,   {Murphy} D.~C.,  2004, \mn@doi [\aj] {10.1086/424934}, \href
  {http://adsabs.harvard.edu/abs/2004AJ....128.2239P} {128, 2239}

\bibitem[\protect\citeauthoryear{{Preston}}{{Preston}}{2011}]{2011AJ....141....6P}
{Preston} G.~W.,  2011, \mn@doi [\aj] {10.1088/0004-6256/141/1/6}, \href
  {http://adsabs.harvard.edu/abs/2011AJ....141....6P} {141, 6}

\bibitem[\protect\citeauthoryear{Price-Whelan}{Price-Whelan}{2017}]{gala}
Price-Whelan A.~M.,  2017, \mn@doi [The Journal of Open Source Software]
  {10.21105/joss.00388}, 2

\bibitem[\protect\citeauthoryear{{Price-Whelan} \& {Johnston}}{{Price-Whelan}
  \& {Johnston}}{2013}]{2013ApJ...778L..12P}
{Price-Whelan} A.~M.,  {Johnston} K.~V.,  2013, \mn@doi [\apjl]
  {10.1088/2041-8205/778/1/L12}, \href
  {http://adsabs.harvard.edu/abs/2013ApJ...778L..12P} {778, L12}

\bibitem[\protect\citeauthoryear{{Price-Whelan}, {Sesar}, {Johnston}  \&
  {Rix}}{{Price-Whelan} et~al.}{2016}]{2016ApJ...824..104P}
{Price-Whelan} A.~M.,  {Sesar} B.,  {Johnston} K.~V.,   {Rix} H.-W.,  2016,
  \mn@doi [\apj] {10.3847/0004-637X/824/2/104}, \href
  {http://adsabs.harvard.edu/abs/2016ApJ...824..104P} {824, 104}

\bibitem[\protect\citeauthoryear{{Rau} et~al.,}{{Rau}
  et~al.}{2009}]{2009PASP..121.1334R}
{Rau} A.,  et~al., 2009, \mn@doi [\pasp] {10.1086/605911}, \href
  {http://adsabs.harvard.edu/abs/2009PASP..121.1334R} {121, 1334}

\bibitem[\protect\citeauthoryear{{Saha}}{{Saha}}{1985}]{1985ApJ...289..310S}
{Saha} A.,  1985, \mn@doi [\apj] {10.1086/162890}, \href
  {http://adsabs.harvard.edu/abs/1985ApJ...289..310S} {289, 310}

\bibitem[\protect\citeauthoryear{{Sales} et~al.,}{{Sales}
  et~al.}{2008}]{2008MNRAS.389.1391S}
{Sales} L.~V.,  et~al., 2008, \mn@doi [\mnras]
  {10.1111/j.1365-2966.2008.13659.x}, \href
  {http://adsabs.harvard.edu/abs/2008MNRAS.389.1391S} {389, 1391}

\bibitem[\protect\citeauthoryear{{Sanders} \& {Binney}}{{Sanders} \&
  {Binney}}{2013}]{2013MNRAS.433.1813S}
{Sanders} J.~L.,  {Binney} J.,  2013, \mn@doi [\mnras] {10.1093/mnras/stt806},
  \href {http://adsabs.harvard.edu/abs/2013MNRAS.433.1813S} {433, 1813}

\bibitem[\protect\citeauthoryear{{Sanderson}}{{Sanderson}}{2016}]{2016ApJ...818...41S}
{Sanderson} R.~E.,  2016, \mn@doi [\apj] {10.3847/0004-637X/818/1/41}, \href
  {http://adsabs.harvard.edu/abs/2016ApJ...818...41S} {818, 41}

\bibitem[\protect\citeauthoryear{{Sanderson}, {Helmi}  \& {Hogg}}{{Sanderson}
  et~al.}{2015}]{2015ApJ...801...98S}
{Sanderson} R.~E.,  {Helmi} A.,   {Hogg} D.~W.,  2015, \mn@doi [\apj]
  {10.1088/0004-637X/801/2/98}, \href
  {http://adsabs.harvard.edu/abs/2015ApJ...801...98S} {801, 98}

\bibitem[\protect\citeauthoryear{{Schlafly} \& {Finkbeiner}}{{Schlafly} \&
  {Finkbeiner}}{2011}]{2011ApJ...737..103S}
{Schlafly} E.~F.,  {Finkbeiner} D.~P.,  2011, \mn@doi [\apj]
  {10.1088/0004-637X/737/2/103}, \href
  {http://adsabs.harvard.edu/abs/2011ApJ...737..103S} {737, 103}

\bibitem[\protect\citeauthoryear{{Sch{\"o}nrich}, {Binney}  \&
  {Dehnen}}{{Sch{\"o}nrich} et~al.}{2010}]{2010MNRAS.403.1829S}
{Sch{\"o}nrich} R.,  {Binney} J.,   {Dehnen} W.,  2010, \mn@doi [\mnras]
  {10.1111/j.1365-2966.2010.16253.x}, \href
  {http://adsabs.harvard.edu/abs/2010MNRAS.403.1829S} {403, 1829}

\bibitem[\protect\citeauthoryear{{Scowcroft}, {Freedman}, {Madore}, {Monson},
  {Persson}, {Seibert}, {Rigby}  \& {Sturch}}{{Scowcroft}
  et~al.}{2011}]{2011ApJ...743...76S}
{Scowcroft} V.,  {Freedman} W.~L.,  {Madore} B.~F.,  {Monson} A.~J.,  {Persson}
  S.~E.,  {Seibert} M.,  {Rigby} J.~R.,   {Sturch} L.,  2011, \mn@doi [\apj]
  {10.1088/0004-637X/743/1/76}, \href
  {http://adsabs.harvard.edu/abs/2011ApJ...743...76S} {743, 76}

\bibitem[\protect\citeauthoryear{{Sesar} et~al.,}{{Sesar}
  et~al.}{2010}]{2010ApJ...708..717S}
{Sesar} B.,  et~al., 2010, \mn@doi [\apj] {10.1088/0004-637X/708/1/717}, \href
  {http://adsabs.harvard.edu/abs/2010ApJ...708..717S} {708, 717}

\bibitem[\protect\citeauthoryear{{Sesar} et~al.,}{{Sesar}
  et~al.}{2013}]{2013ApJ...776...26S}
{Sesar} B.,  et~al., 2013, \mn@doi [\apj] {10.1088/0004-637X/776/1/26}, \href
  {http://adsabs.harvard.edu/abs/2013ApJ...776...26S} {776, 26}

\bibitem[\protect\citeauthoryear{{Sesar} et~al.,}{{Sesar}
  et~al.}{2015}]{2015ApJ...809...59S}
{Sesar} B.,  et~al., 2015, \mn@doi [\apj] {10.1088/0004-637X/809/1/59}, \href
  {http://adsabs.harvard.edu/abs/2015ApJ...809...59S} {809, 59}

\bibitem[\protect\citeauthoryear{{Sesar}, {Hernitschek}, {Dierickx}, {Fardal}
  \& {Rix}}{{Sesar} et~al.}{2017}]{2017ApJ...844L...4S}
{Sesar} B.,  {Hernitschek} N.,  {Dierickx} M.~I.~P.,  {Fardal} M.~A.,   {Rix}
  H.-W.,  2017, \mn@doi [\apjl] {10.3847/2041-8213/aa7c61}, \href
  {http://adsabs.harvard.edu/abs/2017ApJ...844L...4S} {844, L4}

\bibitem[\protect\citeauthoryear{{Sohn} et~al.,}{{Sohn}
  et~al.}{2016}]{2016ApJ...833..235S}
{Sohn} S.~T.,  et~al., 2016, \mn@doi [\apj] {10.3847/1538-4357/833/2/235},
  \href {http://adsabs.harvard.edu/abs/2016ApJ...833..235S} {833, 235}

\bibitem[\protect\citeauthoryear{{Stetson}}{{Stetson}}{1987}]{1987PASP...99..191S}
{Stetson} P.~B.,  1987, \mn@doi [\pasp] {10.1086/131977}, \href
  {http://adsabs.harvard.edu/abs/1987PASP...99..191S} {99, 191}

\bibitem[\protect\citeauthoryear{{Stetson}}{{Stetson}}{1994}]{1994PASP..106..250S}
{Stetson} P.~B.,  1994, \mn@doi [\pasp] {10.1086/133378}, \href
  {http://adsabs.harvard.edu/abs/1994PASP..106..250S} {106, 250}

\bibitem[\protect\citeauthoryear{{Stokes}, {Evans}, {Viggh}, {Shelly}  \&
  {Pearce}}{{Stokes} et~al.}{2000}]{2000Icar..148...21S}
{Stokes} G.~H.,  {Evans} J.~B.,  {Viggh} H.~E.~M.,  {Shelly} F.~C.,   {Pearce}
  E.~C.,  2000, \mn@doi [\icarus] {10.1006/icar.2000.6493}, \href
  {http://adsabs.harvard.edu/abs/2000Icar..148...21S} {148, 21}

\bibitem[\protect\citeauthoryear{{Watkins} et~al.,}{{Watkins}
  et~al.}{2009}]{2009MNRAS.398.1757W}
{Watkins} L.~L.,  et~al., 2009, \mn@doi [\mnras]
  {10.1111/j.1365-2966.2009.15242.x}, \href
  {http://adsabs.harvard.edu/abs/2009MNRAS.398.1757W} {398, 1757}

\bibitem[\protect\citeauthoryear{{Werner} et~al.,}{{Werner}
  et~al.}{2004}]{2004ApJS..154....1W}
{Werner} M.~W.,  et~al., 2004, \mn@doi [\apjs] {10.1086/422992}, \href
  {http://adsabs.harvard.edu/abs/2004ApJS..154....1W} {154, 1}

\bibitem[\protect\citeauthoryear{{White} \& {Rees}}{{White} \&
  {Rees}}{1978}]{1978MNRAS.183..341W}
{White} S.~D.~M.,  {Rees} M.~J.,  1978, \mn@doi [\mnras]
  {10.1093/mnras/183.3.341}, \href
  {http://adsabs.harvard.edu/abs/1978MNRAS.183..341W} {183, 341}

\bibitem[\protect\citeauthoryear{{Xue} et~al.,}{{Xue}
  et~al.}{2008}]{2008ApJ...684.1143X}
{Xue} X.~X.,  et~al., 2008, \mn@doi [\apj] {10.1086/589500}, \href
  {http://adsabs.harvard.edu/abs/2008ApJ...684.1143X} {684, 1143}

\makeatother
\end{thebibliography}



\appendix
\section{SMHASH Light Curves}
\begin{figure*}
\includegraphics[width=\textwidth]{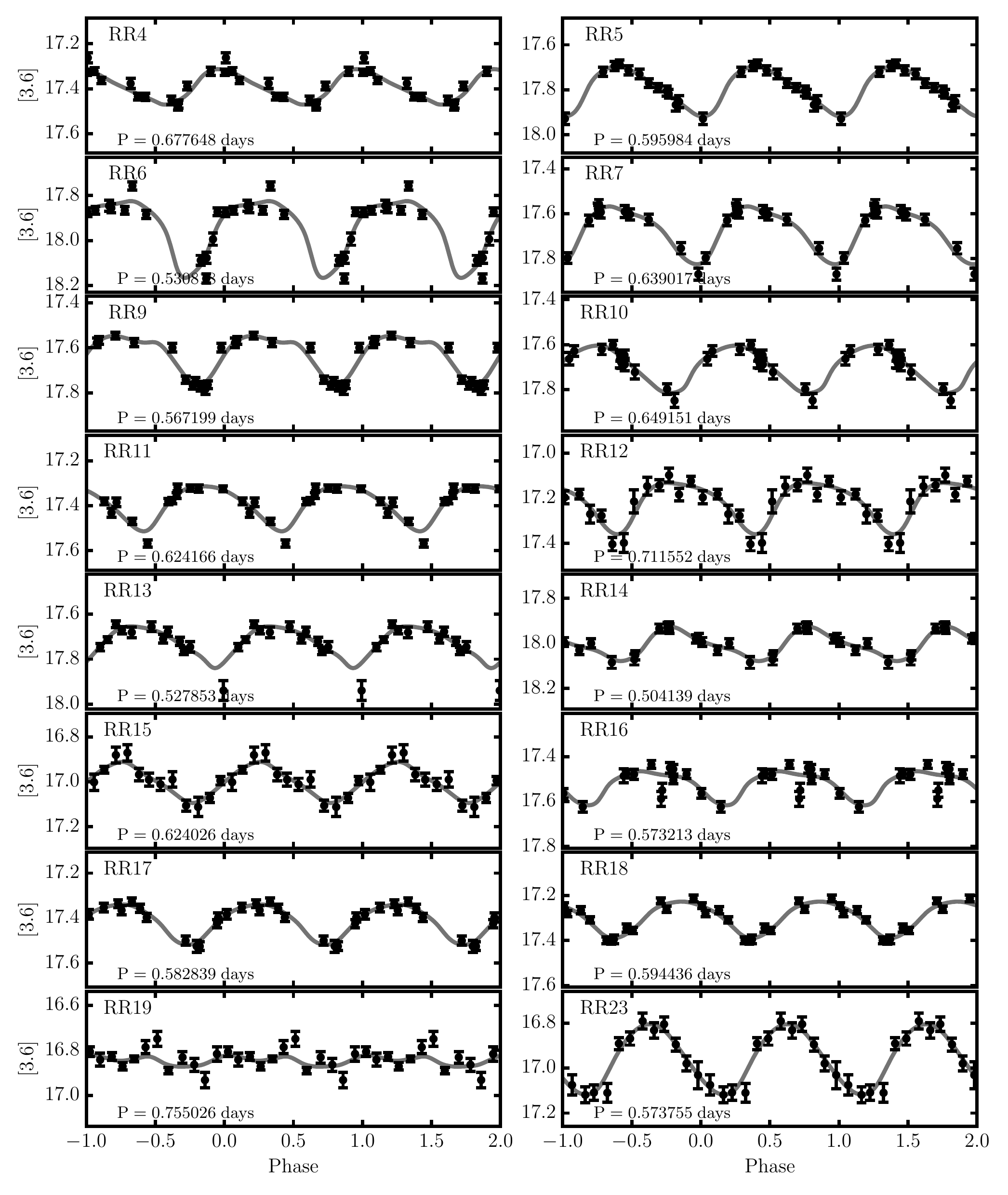}
\caption{{\it Spitzer} 3.6$\mu$m lightcurves of the 32 SMHASH Orphan Stream RR Lyrae star candidates. Each was observed in 12 epochs; the data and fitted lightcurves are repeated through three phase cycles for visual clarity. All stars are shown on the same scale so that amplitude variation is visible. The periods shown were measured from the archival optical survey data \citep{2013ApJ...776...26S}. RR19 is likely not an RR Lyrae star or a stream member but we include it here for completeness. \label{lcs}\label{lc1}}
\end{figure*}

\begin{figure*}\ContinuedFloat
\includegraphics[width=\textwidth]{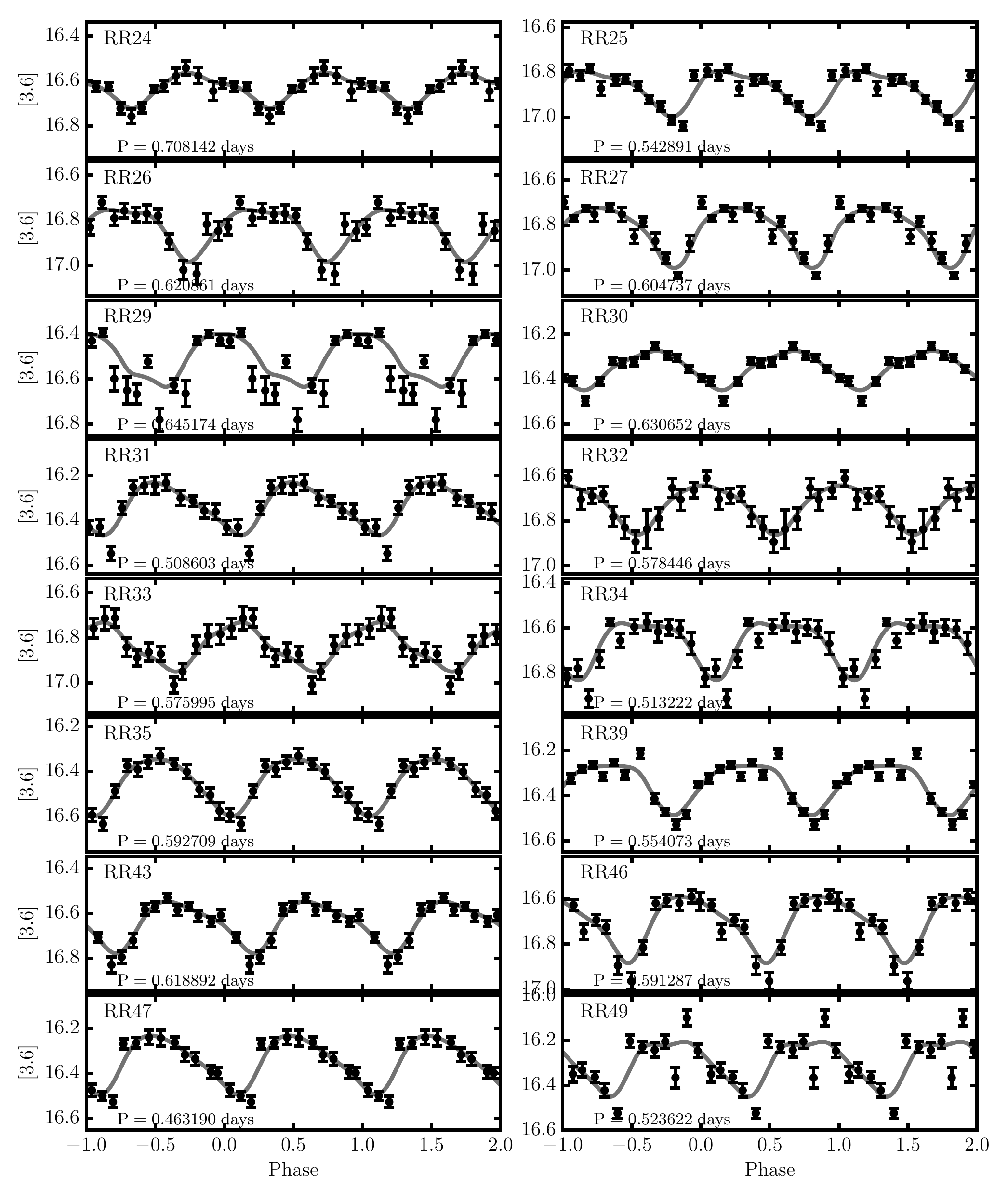}
\caption{Lightcurves, continued}
\end{figure*}



\bsp	
\label{lastpage}
\end{document}